\documentclass[conference]{IEEEtran}
\IEEEoverridecommandlockouts
\usepackage{cite}
\usepackage{amsmath,amssymb,amsfonts}
\usepackage{algorithmic}
\usepackage[ruled,vlined]{algorithm2e}
\usepackage{graphicx}
\usepackage{subfigure}
\usepackage{textcomp}
\usepackage{xcolor}
\def\BibTeX{{\rm B\kern-.05em{\sc i\kern-.025em b}\kern-.08em
    T\kern-.1667em\lower.7ex\hbox{E}\kern-.125emX}}
\begin{document}

\title{Semantic Random Walk for Graph Representation Learning in Attributed Graphs\\
}

\author{\IEEEauthorblockN{Meng Qin}
\IEEEauthorblockA{\textit{Independent Research}\\
mengqin\_az@foxmail.com
}
}

\maketitle

\begin{abstract}
In this study, we focus on the graph representation learning (a.k.a. network embedding) in attributed graphs. Different from existing embedding methods that treat the incorporation of graph structure and semantic as the simple combination of two optimization objectives, we propose a novel semantic graph representation (SGR) method to formulate the joint optimization of the two heterogeneous sources into a common high-order proximity based framework. Concretely, we first construct an auxiliary weighted graph, where the complex homogeneous and heterogeneous relations among nodes and attributes in the original graph are comprehensively encoded. Conventional embedding methods that consider high-order topology proximities can then be easily applied to the newly constructed graph to learn the representations of both node and attribute while capturing the nonlinear high-order intrinsic correlation inside or among graph structure and semantic. The learned attribute embeddings can also effectively support some semantic-oriented inference tasks (e.g., semantic community detection), helping to reveal the graph's deep semantic. The effectiveness of SGR is further verified on a series of real graphs, where it achieves impressive performance over other baselines.
\end{abstract}


\section{Introduction}\label{Sec:Intro}
Graph is an effective model and data structure to describe the entities and relations of various complex systems (e.g., social networks and communication networks). Graph representation learning (a.k.a. network embedding), which aims to encode a graph into a low-dimensional representation with the primary properties preserved, has emerged as an important topic in the research of complex network analysis \cite{Bandy2018FSCNMF}, due to its powerful ability to support the downstream graph inference tasks (e.g., community detection, link prediction, etc.) \cite{Cui2017A}.

As reviewed in \cite{Khosla2019A,Cui2017A}, graph structure (i.e., topology) is a significant information source available for most embedding approaches. For instance, \cite{Tang2015LINE} explored the observed graph topology and latent neighborhood similarity (i.e., the first-order and second-order proximity) respectively in two optimization objectives, while high-order proximities (i.e., neighbor structures with $k$-step random walk) were comprehensively considered in \cite{Perozzi2014DeepWalk,Aditya2016node2vec,Cao2015GraRep}. Besides the microscopic structure (i.e., local neighbor proximity), the mesoscopic community structure can further help to reveal the function and organization of a graph \cite{qin2018adaptive,li2019identifying,qin2021dual,qintowards}. Typical community-preserved representation methods include \cite{Liang2016Modularity,Wang2017Community}.

In this study, we focus on the graph representation learning in attributed graphs, where graph semantic (e.g., node attributes) is another significant heterogeneous information source. It's strongly believed that graph attributes carry orthogonal and complementary knowledge beyond the topology \cite{Jin2018Integrative}, which can potentially enhance learned representations and improve the performance of downstream graph inference tasks. Several embedding methods have been proposed to integrate such two heterogeneous sources, including the matrix factorization (MF) based methods \cite{Cheng2015Network,Bandy2018FSCNMF,Huang2017Accelerated} and deep learning based approaches \cite{Jin2018Integrative,Cao2018Incorporating}. Despite their effectiveness, there remain several limitations.

First, most hybrid methods treat the integration of graph semantic as the auxiliary regularization (to existing models only consider topology) or simply combine the respective optimization objectives to learn features of the two sources, where the nonlinear high-order correlation between graph structure and semantic is not fully explored. For example, if node $v_i$'s neighbors $N(v_i)$ have more shared attributes $\{{a_w}\}$, $v_i$ is more likely to be semantically similar with $N(v_i)$ even though it doesn't directly have all the attributes in $\{{a_w}\}$. Existing methods may fail to capture such high-order correlation between the two heterogeneous sources in a nonlinear manner, which can potentially lead to better performance for the downstream applications.

Moreover, embeddings learned by most existing approaches with topology and attribute may only be capable to improve the performance of some simple inference tasks (e.g., community detection), but cannot be directly applied to some advanced semantic-oriented downstream tasks, e.g., semantic community detection \cite{Wang2016Semantic,qin2018adaptive,li2019identifying,qin2019towards,qin2021dual}, where one can obtain the corresponding semantic descriptions of each community simultaneously when the community partition is finished. For most existing embedding methods, additional efforts still need to be taken after the basic inference (e.g., community partition) to generate such descriptions (in order to explore graph semantic) with the undesired loss of correlations between the two sources.

We introduce a novel semantic graph representation (SGR) model to alleviate the aforementioned limitations, in which both the node and attribute (in the original graph $G$) are treated as entities in an auxiliary weighted graph $G'$. In particular, $G'$ comprehensively encodes 3 types of relations (i.e., relations between (\romannumeral1) node pairs $\{({v_i},{v_j})\}$, (\romannumeral2) attribute pairs $\{({a_w},{a_s})\}$ and (\romannumeral3) heterogeneous entity pairs $\{({v_i},{a_w})\}$). In this case, conventional high-order proximity based embedding methods (e.g., DeepWalk \cite{Perozzi2014DeepWalk}) can be easily applied to $G'$ to jointly learn the low-dimensional representations of both nodes $\{{v_i}\}$ and attributes $\{{a_w}\}$, where the nonlinear high-order intrinsic correlations among graph topology and attribute are fully captured. Besides the node representations, the attribute embeddings learned by SGR can also be effectively used to support the semantic-oriented inference (e.g., semantic community detection), revealing the deep semantic of a graph.

We summarize our main contributions as follows. \textbf{(\romannumeral1)} We formulate the network embedding in attributed graphs as the embedding task of an auxiliary weighted graph with heterogeneous entities, so that conventional high-order proximity based methods can be used to explore the graph semantic. \textbf{(\romannumeral2)} We proposed a novel SGR method, which can not only fully utilize the attribute information to improve the embedding quality but can also generate the semantic descriptions to support the advanced semantic-oriented graph inference tasks. \textbf{(\romannumeral3)} An enhancement scheme based on graph regularization is also introduced for SGR to explore the effect of other side information (e.g., community structure). \textbf{(\romannumeral4)} To verify the effectiveness of SGR, we conduct extensive experiments on a series of real graphs, where SGR consistently outperforms other baselines.

In the rest of this paper, we first give the formal problem statements regarding network embedding in Section~\ref{Sec:Prob} and elaborate on the proposed SGR method in Section~\ref{Sec:Meth}. Experiments are described in Section~\ref{Sec:Exp}, including the performance evaluation on real graph datasets and a case study about semantic descriptions. Section~\ref{Sec:Conc} concludes this paper and indicates our future work.

\section{Problem Statements}\label{Sec:Prob}
In this study, we consider the graph representation learning (a.k.a. network embedding) of undirected graphs with discrete node attributes. Assume that there are $n$ nodes and $e$ edges in a graph as well as $m$ discrete attributes associated with each node. The graph can be described as a 4-tuple $G = (V,E,A,F)$, where $V = \{ {v_1},\cdots ,{v_n}\}$ is the set of nodes; $E = \{ ({v_i},{v_j})\left| {{v_i},{v_j} \in V} \right.\}$ is the set of edges; $A = \{ {a_1},\cdots ,{a_m}\}$ is the set of attributes; $F = \{ f({v_1}),\cdots ,f({v_n})\}$ denotes the map from $V$ to $A$, with $f({v_i}) \subset A$ as the set of attributes of ${v_i}$.

Given $G$, the goal of network embedding is to learn a function $f:\{ {v_i}\}  \mapsto \{ {{\bf{x}}_i} \in {\Re ^{1 \times k}}\}$ that maps each node $v_i$ to a $k$-dimensional vector ${{\bf{x}}_i}$ (with $k \ll \min \{ n,m\}$), in which the major properties of graph topology and semantic (hidden in $E$ and $F$) are comprehensively preserved. Namely, the node pair $({v_i},{v_j})$ with similar properties (e.g., community membership and semantic) should have similar representations $({{\bf{x}}_i},{{\bf{x}}_j})$.

In particular, we formulate the aforementioned representation learning procedure as the embedding task of an auxiliary weighted graph ${G'}$, where each node ${v_i}$ and attribute ${a_j}$ in original graph $G$ are abstracted as the heterogeneous nodes. Concretely, we use a 2-tuple $G' = ( V',E' )$ to described the weighted graph, where $V' = V \cup A$ is the set of heterogeneous nodes and $E' = \{ {E_1},{E_2},{E_3}\}$ is the set of relations (i.e., heterogeneous edges), with ${E_1} = \{ W({v_i},{v_j})|{v_i},{v_j} \in V\}$, ${E_2} = \{ W({v_i},{a_w})|{v_i} \in V,{a_w} \in A\}$ and ${E_3} = \{ W({a_w},{a_s})|{a_w},{a_s} \in A\}$ as the set of weighted edges between (original) node pairs $\{({v_i},{v_j})\}$, heterogeneous entity pairs $\{({v_i},{a_w})\}$ and attribute pairs $\{({a_w},{a_s})\}$, respectively. Based on ${G'}$, the goal of SGR is to learn a function $f':\{ {v_i},{a_j}\}  \mapsto \{ {{{\bf{x'}}}_{{v_i}}},{{{\bf{x'}}}_{{a_j}}} \in \Re {}^{1 \times k}\}$ that maps each node/attribute ${v_i}$/${a_j}$ into a $k$-dimensional hidden space represented by vector ${{{\bf{x'}}}_{{v_i}}}$/${{{\bf{x'}}}_{{a_j}}}$, in which the key properties of ${G'}$ are comprehensively encoded. Given the learned representations $\{{{{\bf{x'}}}_{{v_i}}},{{{\bf{x'}}}_{{a_i}}}\}$, one can treat $\{{{{\bf{x'}}}_{{v_i}}}\}$ as the final result (i.e., node vectors) of the embedding task. Furthermore, $\{{{{\bf{x'}}}_{{a_i}}}\}$ can also be utilized to generate the semantic descriptions for each node or (node) cluster by selecting the top nearest attribute entities of the specific node or cluster center in the mapped hidden space. 

\section{Methodology}\label{Sec:Meth}
\begin{figure*}[!t]
\centering
\includegraphics[width=0.55\linewidth]{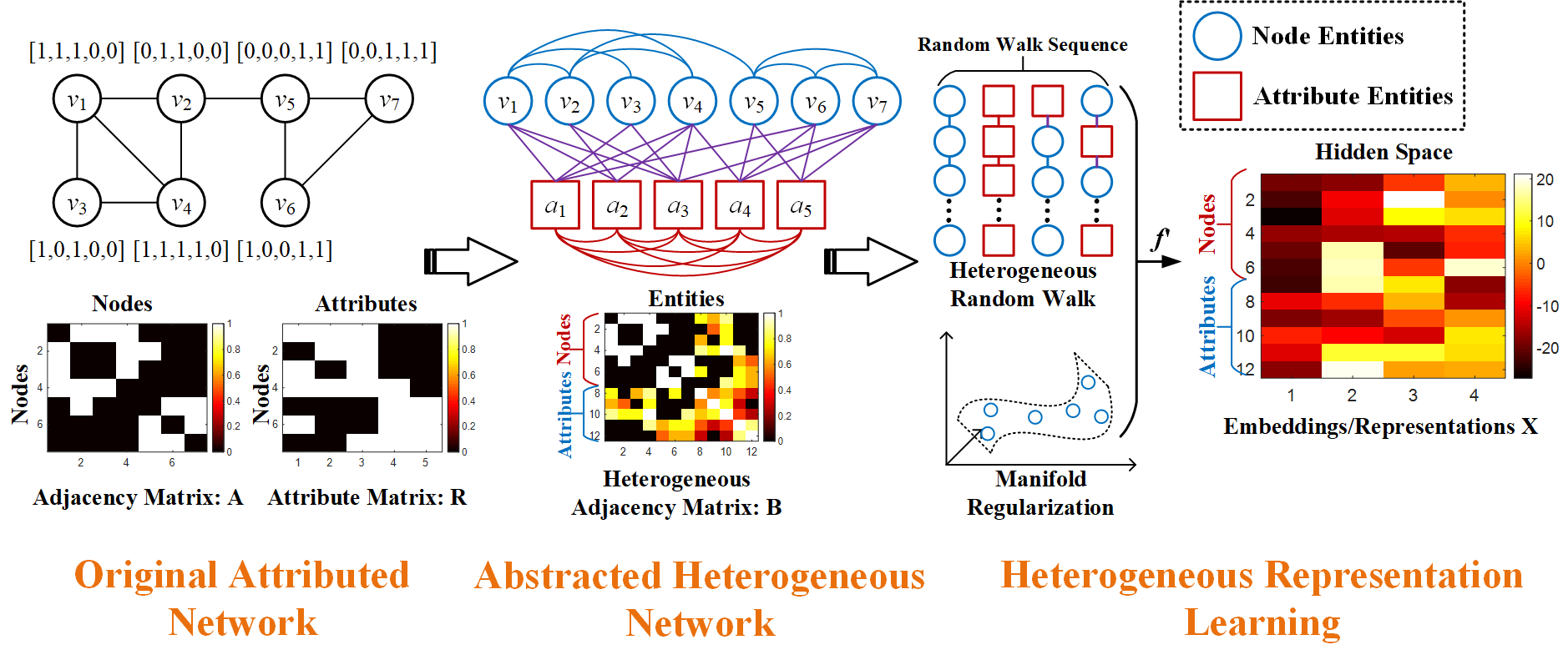}
\caption{Overview of our SGR method, where we first (\romannumeral1) construct an auxiliary weighted graph according to original graph's topology as well as attributes and (\romannumeral2) use high-order proximity based embedding methods and manifold regularization techniques to learn the embeddings of both nodes and attributes.}\label{SGR-Sketch}
\end{figure*}

We propose a novel semantic graph representation (SGR) method to tackle the embedding task of attributed graphs. A high-level overview of SGR is presented in Fig.~\ref{SGR-Sketch}.

As illustrated in Fig.~\ref{SGR-Sketch}, SGR integrates three types of relations (i.e., node relations ${E_1}$, heterogeneous relations ${E_2}$, and attribute relations ${E_3}$) into a unified weighted graph $G'$, where the topology and attributes of the original attributed graph $G$ are comprehensively encoded in $G'$'s weighted topology. The deep knowledge of $G'$ is then fully explored and embedded in a low-dimensional hidden space by utilizing some random walk based embedding methods and graph regularization techniques on $G'$. Details of SGR are elaborated in the rest of this section.

\textbf{Modeling Node Relations.} The node relations ${E_1}$ in $G'$ describe the topology structure of $G$. They can be described by an \textit{adjacency matrix} ${\bf{A}} \in {\Re ^{n \times n}}$, where ${{\bf{A}}_{ij}} = {{\bf{A}}_{ji}} = 1$ if there is an edge between node pair $({v_i},{v_j})$ and ${{\bf{A}}_{ij}} = {{\bf{A}}_{ji}} = 0$, otherwise.
%For SGR, we utilize ${\bf{A}}$ to represent the homogeneous relations between the node entities $\{{v_i} \in V\}$ (i.e., node relations ${E_1}$) in $G'$.

\textbf{Modeling Attribute Relations.} The attribute relations ${E_3}$ (i.e., homogeneous relations between attribute entities $\{{a_w} \in A\}$) in $G'$ describe the similarity between each attribute pair $({a_s},{a_w})$. In general, the graph semantic of $G$ can be described by a \textit{node attribute matrix} ${{\bf{R}}_0} \in {\Re ^{n \times m}}$, where ${({{\bf{R}}_0})_{iw}} = 1$ if node ${v_i}$'s attribute set has ${a_w}$ and  ${({{\bf{R}}_0})_{iw}} = 0$, otherwise. ${({{\bf{R}}_0})_{iw}}$ can also be defined as the occurrence frequency or IF/IDF value of $a_w$ in $v_i$'s attribute set. In particular, ${({{\bf{R}}_0})_{:,w}}$ describes the node membership of a certain attribute ${a_w}$ (i.e., which nodes have $a_w$ in their attribute sets). Hence, we define the similarity between each attribute pair $({a_s},{a_w})$ as the normalized similarity between their node memberships $( {({{\bf{R}}_0})_{:w}},{({{\bf{R}}_0})_{:s}})$ and introduce the \textit{node similarity matrix} ${\bf{P}} \in {\Re ^{m \times m}}$, where
\begin{equation}
    {\bf{P}} = {{\bf{D}}^{{{ - 1} \mathord{\left/
    {\vphantom {{ - 1} 2}} \right.
    \kern-\nulldelimiterspace} 2}}}{{\bf{P}}_0}{{\bf{D}}^{ - {1 \mathord{\left/
    {\vphantom {1 2}} \right.
    \kern-\nulldelimiterspace} 2}}},
\end{equation}
with ${({{\bf{P}}_0})_{ws}} = {{[({{\bf{R}}_0})_{:,w}^T{{({{\bf{R}}_0})}_{:,s}}]} \mathord{\left/ {\vphantom {{[({{\bf{R}}_0})_{:,w}^T{{({{\bf{R}}_0})}_{:,s}}]} {[|{{({{\bf{R}}_0})}_{:,w}}| \cdot |{{({{\bf{R}}_0})}_{:,s}}|]}}} \right. \kern-\nulldelimiterspace} {[|{{({{\bf{R}}_0})}_{:,w}}| \cdot |{{({{\bf{R}}_0})}_{:,s}}|]}}$ and ${\bf{D}} = {\mathop{\rm diag}\nolimits} (\sum\nolimits_{s = 1}^m {{{({{\bf{P}}_0})}_{1s}}} , \cdots ,\sum\nolimits_{s = 1}^m {{{({{\bf{P}}_0})}_{ms}}} )$.

Note that there may exist the magnitude difference between ${\bf{A}}$ and ${\bf{P}}$ when integrating them into $G'$, unfairly affecting the learned the heterogeneous embeddings. To eliminate the difference, we use the Max-Min normalization to rescale the elements in ${\bf{P}}$ into the range $[0,1]$ (with the result notated as ${{\bf{\tilde P}}}$). Given a specific vector/matrix input $\bf{s}$ (e.g., ${\bf{P}}$), the Max-Min normalization (notated as MNorm) is defined as follow:
\begin{equation}
    {{\bf{\tilde s}}} = {\mathop{\rm MNorm}\nolimits} ({{\bf{s}}_i}) = {{({{\bf{s}}_i} - {{\bf{s}}_{\min }})} \mathord{\left/
    {\vphantom {{({{\bf{s}}_i} - {{\bf{s}}_{\min }})} {({{\bf{s}}_{\max }} - {{\bf{s}}_{\min }})}}} \right.
    \kern-\nulldelimiterspace} {({{\bf{s}}_{\max }} - {{\bf{s}}_{\min }})}},
\end{equation}
where ${{\bf{s}}_{\min }}$ and ${{\bf{s}}_{\max }}$ are the minimum and maximum elements in $\bf{s}$, respectively.

\begin{figure}[!t]
\centering
\includegraphics[width=0.55\linewidth]{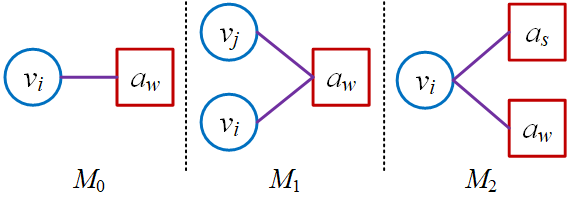}
\caption{The heterogeneous entity motifs considered in this study with circle and square denoting node and attribute entities, respectively.} \label{Motif-Fig}
\end{figure}

\textbf{Modeling Heterogeneous Relations.} We adopt motif, a substructure that reveals the high-order organization and function of a graph, to describe the heterogeneous relations $E_2$ between each heterogeneous entity pair $({v_i},{a_w})$. As a demonstration, we consider three basic motif instances ${M_0} = \{ ({v_i},{a_w}) |{v_i} \in V,{a_w} \in A\}$, ${M_1} = \{ ({v_i},{a_w}),({v_j},{a_w}) |{v_i},{v_j} \in V,{a_w} \in A\}$ and ${M_2} = \{ ({v_i},{a_w}),({v_i},{a_s}) |{v_i} \in V,{a_w},{a_s} \in A\}$, which are illustrated in Fig.~\ref{Motif-Fig} with circle and square denoting node and attribute entities, respectively.

Concretely, ${M_0}$, which is described by a \textit{relation matrix} ${\bf{R}}_0 \in \Re^{n \times m}$, encodes the observable relations while ${M_1}$ and ${M_2}$ encode the higher-order structures that preserve the deep knowledge of heterogeneous relations (e.g., nodes shared with more attributes should be more similar in their properties). To further describe the higher-order structures of ${M_1}$ and ${M_2}$, we introduce another two \textit{relation matrices} ${{\bf{R}}_1} \in {\Re ^{n \times m}}$ and ${{\bf{R}}_2} \in {\Re ^{n \times m}}$, where ${({{\bf{R}}_t})_{iw}}$ is the co-occurrence counts of node ${v_i}$ and attribute ${a_w}$ in motif ${M_t}$ ($t \in \{ 1,2\}$) (i.e., the number of stances ${M_t}$ containing $({v_i},{a_w})$) or the cumulative sum of $({v_i},{a_w})$'s weight (i.e., ${({{\bf{R}}_0})_{iw}}$) in ${M_t}$. In this case, ${{\bf{R}}_1}$ and ${{\bf{R}}_2}$ can be considered as the rescaling of ${{\bf{R}}_0}$'s observed relations, where relation $({v_i},{a_w})$ may have large weight ${({{\bf{R}}_t})_{iw}}$ if it can reflect the major property of ${M_t}$.

We further conduct normalization on $\{ {{\bf{R}}_0},{{\bf{R}}_1},{{\bf{R}}_2}\}$ (with results notated as $\{ {{{\bf{\tilde R}}}_0},{{{\bf{\tilde R}}}_1},{{{\bf{\tilde R}}}_2}\}$), since there remains magnitude difference among them. Moreover, we use the combination of $\{ {{{\bf{\tilde R}}}_0},{{{\bf{\tilde R}}}_1},{{{\bf{\tilde R}}}_2}\}$ to consider the comprehensive effect of different motifs by setting
\begin{equation}
    {\bf{R}} = {\delta _0}{{{\bf{\tilde R}}}_0} + {\delta _1}{{{\bf{\tilde R}}}_1} + {\delta _2}{{{\bf{\tilde R}}}_2},
\end{equation}
with $\{{\delta _0},{\delta _1},{\delta _2}\}$ as parameters to control different components. We let ${\delta _0}={\delta _1}={\delta _2}=1$ as the default setting, where $\{ {M_0},{M_1},{M_2}\}$ have the same contribution in (3) without specific prior knowledge. Better performance can be achieved by fine-tuning ${\delta _0},{\delta _1},{\delta _2} \in \{ 0,1\}$ according to our experiments (see Section 4.1). Similar to ${\bf{P}}$, we also conducted another normalization on ${\bf{R}}$ to further eliminate the magnitude difference between ${\bf{R}}$ and $\{ {\bf{A}},{\bf{\tilde P}}\}$, with the result notated as ${{\bf{\tilde R}}}$.

\textbf{The Unified Model.} We further construct an auxiliary graph $G'$ with its weighted topology described by the following \textit{heterogeneous adjacency matrix} ${\bf{B}} \in {\Re ^{(n + m) \times (n + m)}}$:
\begin{equation}
    {\bf{B}} = \left[ {\begin{array}{*{20}{c}}
    {\bf{A}}&{{\bf{\tilde R}}}\\
    {{{{\bf{\tilde R}}}^T}}&{{\bf{\tilde P}}}
    \end{array}} \right],
\end{equation}
where $\{ {\bf{A}},{\bf{\tilde P}},{\bf{\tilde R}}\}$ are considered as different blocks of ${\bf{B}}$, and ${{\bf{B}}_{ij}}$ is the edge weight of entity pair $({e_i},{e_j})$ (${e_i},{e_j} \in V \cup A$). In this case, existing high-order proximity based embedding methods can be easily applied to $G'$ to fully explore both the information sources of graph structure and semantic.

As a demonstration, we adopt the following matrix factorization (MF) objective of DeepWalk \cite{Qiu2018Network} as an example to derive the heterogeneous embeddings:
\begin{equation}
    {\bf{Z}} = \log ({\mathop{\rm vol}\nolimits} (G') \cdot (\frac{1}{o}\sum\nolimits_{r = 1}^o {{{({{\bf{D}}^{ - 1}}{\bf{B}})}^r}} ){{\bf{D}}^{ - 1}}) - \log b,
\end{equation}
where $o$ is the content window size (i.e, step/order of random walk); $b$ is the number of negative sampling; ${\bf{D}} = {\mathop{\rm diag}\nolimits} ({d_1},{d_2}, \cdots ,{d_{n + m}})$ is the degree diagonal matrix with ${d_i} = \sum\nolimits_{j = 1}^{n + m} {{{\bf{B}}_{ij}}}$ as the degree of entity ${e_i}$; ${\mathop{\rm vol}\nolimits} (G') = \sum\nolimits_{i = 1}^{n + m} {{d_i}}$ is the volume of $G'$. Based on ${\bf{Z}}$, the network embedding task can then be represented as the following MF-based optimization problem:
\begin{equation}
    \mathop {\arg \min }\limits_{{\bf{X}},{\bf{Y}}} \left\| {{\bf{Z}} - {\bf{X}}{{\bf{Y}}^T}} \right\|_F^2,
\end{equation}
where ${\bf{X}} \in {\Re ^{(n + m) \times k}}$ and ${\bf{Y}} \in {R^{(n + m) \times k}}$ are two low-dimensional matrices. The singular value decomposition (SVD) can be utilized to get the optimal solution of (6), which is defined as follow:
\begin{equation}
    {\bf{Z}} = {\bf{U\Sigma }}{{\bf{V}}^T} \approx {{\bf{U}}_{:,1:k}}{{\bf{\Sigma }}_k}{\bf{V}}_{:,1:k}^T.
\end{equation}
In (7), ${\bf{\Sigma }} = diag({\theta _1},{\theta _2}, \cdots ,{\theta _{n + m}})$ is the diagonal matrix of singular values with ${\theta _1} \ge {\theta _2} \ge  \cdots  \ge {\theta _{n + m}}$. We use the top-$k$ singular values to approximatively reconstruct $\bf{Z}$, so the solution of (6) can be derived by setting
\begin{equation}
    {{\bf{X}}^*} = {{\bf{U}}_{:,1:k}}\sqrt {{{\bf{\Sigma }}_k}} ,\mathop {}\limits_{} {{\bf{Y}}^*} = {{\bf{V}}_{:,1:k}}\sqrt {{{\bf{\Sigma }}_k}},
\end{equation}
where ${{\bf{X}}^*}$ is adopted as the final embedding result of SGR. Algorithm~\ref{Alg:SGR} summarizes the aforementioned procedure.

\begin{algorithm}[tb]\scriptsize
\caption{Semantic Graph Representation (SGR)}\label{Alg:SGR}
\LinesNumbered
\KwIn{${\bf{A}}$, ${{\bf{R}}_0}$}
\KwOut{$\{ {{\bf{X}}^*},{{\bf{Y}}^*}\}$}
construct \textit{node similarity matrix} ${\bf{P}}$ via (1)\\
normalize ${\bf{P}}$ via (2) (with the result notated as ${{\bf{\tilde P}}}$)\\
construct \textit{relation matrices} $\{ {{\bf{R}}_1},{{\bf{R}}_2}\}$ according to motifs $\{ {M_1},{M_2}\}$\\
normalize $\{ {{\bf{R}}_0},{{\bf{R}}_1},{{\bf{R}}_2}\}$ via (2) (with the result notated as $\{ {{{\bf{\tilde R}}}_0},{{{\bf{\tilde R}}}_1},{{{\bf{\tilde R}}}_2}\}$)\\
construct \textit{relation matrix} ${\bf{R}}$ via (3)\\
normalize ${\bf{R}}$ via (2) (with the result notated as ${{\bf{\tilde R}}}$)\\
construct \textit{heterogeneous adjacency matrix} ${\bf{B}}$ via (4)\\
construct the optimization objective (5)\\
get the solution of (6) (i.e., $\{ {{\bf{X}}^*},{{\bf{Y}}^*}\}$) by using SVD (i.e., (8))\\
\end{algorithm}

\textbf{Enhancement of Side Information.} In addition to the obserable graph topology and attribute described by $\{ {\bf{A}},{\bf{\tilde R}}\}$, some other latent side information (e.g., community structure) can also be utilized to further enhance the embedding learning. We define such an effect as \textit{side-enhancement}.

We use graph regularization to leverage the side information based on (6). For a certain information source (notated as ${I_l}$), the corresponding regularization term is defined as follow:
\begin{equation}
\resizebox{.91\linewidth}{!}{$
    {{\mathop{\rm Reg}\nolimits} _l}({\bf{X}},{{\bf{T}}_l}) = \frac{1}{2}\sum\limits_{i,j = 1}^n {{{({{\bf{T}}_l})}_{ij}}\left\| {{{\bf{X}}_{i,:}} - {{\bf{X}}_{j,:}}} \right\|_2^2}  = {\mathop{\rm tr}\nolimits} ({{\bf{X}}^T}{{\bf{L}}_l}{\bf{X}}),
$}
\end{equation}
where ${{\bf{T}}_l} \in {\Re ^{(n + m) \times (n + m)}}$ is the matrix encoding the primary properties of ${I_l}$ and ${{\bf{L}}_l} = ({{\bf{D}}_l} - {{\bf{T}}_l})$ is the Laplacian matrix with ${{\bf{D}}_l} = {\mathop{\rm diag}\nolimits} (\sum\limits_{j = 1}^{n + m} {{{({T_l})}_{1j}}} ,\sum\limits_{j = 1}^{n + m} {{{({T_l})}_{2j}}} , \cdots )$. In other words, (9) can be considered as the penalty given by ${I_l}$, in which $\{ {{\bf{X}}_{i,:}},{{\bf{X}}_{j,:}}\}$ are regularized to have similar representations if ${({{\bf{T}}_l})_{ij}}$ has a relatively large value.
As a demonstration, \textit{community structure} and \textit{attribute similarity} are adopted as two available sources of side information.

We utilize the \textit{modularity matrix} ${\bf{Q}} \in {\Re ^{n \times n}}$ \cite{Jin2018Integrative} to encode the community structure of $G$, where
\begin{equation}
    {{\bf{Q}}_{ij}} = {{\bf{Q}}_{ji}} = {{\rm A}_{ij}} - {{{d_i}{d_j}} \mathord{\left/ {\vphantom {{{d_i}{d_j}} {(2e)}}} \right. \kern-\nulldelimiterspace} {(2e)}},
\end{equation}
with ${d_i} = \sum\nolimits_{j = 1}^n {{{\bf{A}}_{ij}}}$ as the degree of node ${v_i}$ and $e$ as the number of edges in $G$. In particular, ${\bf{Q}}$ encodes the primary properties of $G$'s community structure by measuring the difference between the exact edge numbers and the expected number of such edges over all node pairs. ${{\bf{Q}}_{ij}} = {{\bf{Q}}_{ji}}$ with larger values indicates that edge $({v_i},{v_j})$ is more likely to be preserved in a certain community (but not to be cut) when conducting the graph-cut partitioning.

Furthermore, we utilize the cosine similarity between the attribute lists of each node pair $({v_i},{v_j})$ to describe the \textit{attribute similarity} of $G$, where we introduce the \textit{attribute similarity matrix} ${\bf{S}} \in {R^{n \times n}}$ with 
\begin{equation}
\resizebox{.89\linewidth}{!}{$
    {{\bf{S}}_{ij}} = {{\bf{S}}_{ji}} = {{[{{({{\bf{R}}_0})}_{i,:}}({{\bf{R}}_0})_{j,:}^T]} \mathord{\left/
    {\vphantom {{[{{({{\bf{R}}_0})}_{i,:}}({{\bf{R}}_0})_{j,:}^T]} {[|{{({{\bf{R}}_0})}_{i,:}}| \cdot |{{({{\bf{R}}_0})}_{j,:}}|]}}} \right.
    \kern-\nulldelimiterspace} {[|{{({{\bf{R}}_0})}_{i,:}}| \cdot |{{({{\bf{R}}_0})}_{j,:}}|]}}.
$}
\end{equation}

The overall optimization objective of side-enhancement can then be formulated as follow:
\begin{equation}
\resizebox{.89\linewidth}{!}{$
    \mathop {\arg \min }\limits_{{\bf{X}},{\bf{Y}}} {\mathop{\rm O}\nolimits} ({\bf{X}},{\bf{Y}}) = \left\| {{\bf{Z}} - {\bf{X}}{{\bf{Y}}^T}} \right\|_F^2 + \sum\limits_{l = 1}^L {{\lambda _l}{{{\mathop{\rm Reg}\nolimits} }_l}({\bf{X}},{{\bf{T}}_l})},
$}
\end{equation}
where ${\lambda _l}$ is the parameter to adjust the effect of the $l$-th side information. In this study, we set $L=2$ and let
\begin{equation}
    {{\bf{T}}_1} = \left[ {\begin{array}{*{20}{c}}
    {{\bf{\tilde Q}}}&{\bf{0}}\\
    {\bf{0}}&{\bf{0}}
    \end{array}} \right],{{\bf{T}}_2} = \left[ {\begin{array}{*{20}{c}}
    {{\bf{\tilde S}}}&{\bf{0}}\\
    {\bf{0}}&{\bf{0}}
    \end{array}} \right],
\end{equation}
with $\{ {\bf{\tilde S}},{\bf{\tilde Q}}\}$ as the normalized results of $\{ {\bf{S}},{\bf{Q}}\}$ via (4).

To obtain the solution of (12) (notated as $\{ {{{\bf{X'}}}^*},{{{\bf{Y'}}}^*}\}$) in a relatively fast way, we first use (6)'s result (i.e., (8)) to initialize $\{ {\bf{X}},{\bf{Y}}\}$ and use certain rules to update their values.
For ${\bf{X}}$, we derive the partial derivative of ${\mathop{\rm O}\nolimits} ({\bf{X}},{\bf{Y}})$ w.r.t. ${\bf{X}}$:
\begin{equation}
    {{\partial {\mathop{\rm O}\nolimits} ({\bf{X}},{\bf{Y}})} \mathord{\left/
    {\vphantom {{\partial {\mathop{\rm O}\nolimits} ({\bf{X}},{\bf{Y}})} {\partial {\bf{X}}}}} \right.
    \kern-\nulldelimiterspace} {\partial {\bf{X}}}} = 2({\bf{X}}{{\bf{Y}}^T}{\bf{Y}} - {\bf{ZY}} + {\bf{LX}}),
\end{equation}
with ${\bf{L}} = \sum\limits_{l = 1}^L {{\lambda _l}{{\bf{L}}_l}}$. By setting ${{\partial {\mathop{\rm O}\nolimits} ({\bf{X}},{\bf{Y}})} \mathord{\left/ {\vphantom {{\partial {\mathop{\rm O}\nolimits} ({\bf{X}},{\bf{Y}})} {\partial {\bf{X}}}}} \right. \kern-\nulldelimiterspace} {\partial {\bf{X}}}}=0$, we have
\begin{equation}
    {{{\bf{X'}}}^*} = {({{\bf{I}}_n} + {\bf{L}})^\dag }{\bf{ZY}}{({{\bf{Y}}^T}{\bf{Y}} + {{\bf{I}}_k})^\dag },
\end{equation}
in which ${{\bf{M}}^\dag }$ denotes the pseudo-inverse of matrix $\bf{M}$, while ${{\bf{I}}_n}$ denotes an $n$-dimensional identity matrix. 
For $\bf{Y}$, we further derive ${{\mathop{\rm O}\nolimits} ({\bf{X}},{\bf{Y}})}$'s partial derivative w.r.t. $\bf{Y}$:
\begin{equation}
    {{\partial {\mathop{\rm O}\nolimits} ({\bf{X}},{\bf{Y}})} \mathord{\left/
    {\vphantom {{\partial {\mathop{\rm O}\nolimits} ({\bf{X}},{\bf{Y}})} {\partial {\bf{Y}}}}} \right.
    \kern-\nulldelimiterspace} {\partial {\bf{Y}}}} = 2({\bf{Y}}{{\bf{X}}^T}{\bf{X}} - {{\bf{Z}}^T}{\bf{X}}).
\end{equation}
By setting ${{\partial {\mathop{\rm O}\nolimits} ({\bf{X}},{\bf{Y}})} \mathord{\left/ {\vphantom {{\partial {\mathop{\rm O}\nolimits} ({\bf{X}},{\bf{Y}})} {\partial {\bf{Y}}}}} \right. \kern-\nulldelimiterspace} {\partial {\bf{Y}}}} = 0$, we obtain ${\bf{Y}}$'s updating rule:
\begin{equation}
    {{{\bf{Y'}}}^*} = ({{\bf{Z}}^T}{\bf{X}}){({{\bf{X}}^T}{\bf{X}})^\dag }.
\end{equation}

The solution $\{{{{\bf{X'}}}^*},{{{\bf{Y'}}}^*}\}$ can be obtained by iteratively update $\{ {\bf{X}},{\bf{Y}}\}$ via (15) and (17) until converge. According to our pre-experiments, the relative error of objective function (12) (w.r.t. previous iteration) is less than ${10^{ - 6}}$ just after the first iteration, and (14)'s value keep stables in the subsequent iterations on most real graph datasets. Hence, we just use (15) and (17) to update $\{ {\bf{X}},{\bf{Y}}\}$ once after initialization to get the solution in a fast way. Similar to (8), we adopt ${{{\bf{X'}}}^*}$ as the final embedding result. As a conclusion, Algorithm~\ref{Alg:SGR-Side} summarizes the overall procedure of side-enhancement.

\begin{algorithm}[tb]\scriptsize
\caption{Side-Enhancement of SGR}\label{Alg:SGR-Side}
\LinesNumbered
\KwIn{${\bf{A}}$, ${{\bf{R}}_0}$, $\{ {{\bf{X}}^*},{{\bf{Y}}^*}\}$}
\KwOut{$\{{{{\bf{X'}}}^*},{{{\bf{Y'}}}^*}\}$}
construct the \textit{modularity matrix} ${\bf{Q}}$ via (10)\\
normalize ${\bf{Q}}$ via (2) (with the result notated as ${{\bf{\tilde Q}}}$)\\
construct the \textit{attribute similarity matrix} ${\bf{S}}$ via (11)\\
normalize ${\bf{S'}}$ via (2) (with the result notated as ${{\bf{\tilde S'}}}$)\\
construct $\{ {{\bf{T}}_1},{{\bf{T}}_2}\}$ via (13) (according to $\{ {\bf{\tilde S'}},{\bf{\tilde Q}}\}$)\\
use the result of (6) (i.e., (8)) to initialize $\{ {\bf{X}},{\bf{Y}}\}$\\
update ${\bf{X}}$ via (15) (with the result notated as ${{{\bf{X'}}}^*}$)\\
update ${\bf{Y}}$ via (17) (with the result notated as ${{{\bf{Y'}}}^*}$)\\
\end{algorithm}

\section{Experiments}\label{Sec:Exp}
\subsection{Performance Evaluation on Real Graphs}
\textbf{Datasets.} To verify the effectiveness of SGR, we applied it to 12 real attributed graphs. Statistics details of the datasets after necessary pre-processing are shown in Table~\ref{Data-Table}, where $n$, $e$, $m$ and $c$ are the number of nodes, edges, (node) attributes and clusters/classes, respectively.

\begin{table}[]\scriptsize
\caption{Statistics details of the real attributed graphs.}\label{Data-Table}
\begin{tabular}{p{1.2cm}|p{0.1cm}p{0.35cm}p{0.25cm}p{0.05cm}|p{1.15cm}|p{0.25cm}p{0.45cm}p{0.35cm}l}
\hline
\textbf{Datasets} & \textit{\textbf{N}} & \textit{\textbf{E}} & \textit{\textbf{M}} & \textit{\textbf{C}} & \textbf{Datasets} & \textit{\textbf{N}} & \textit{\textbf{E}} & \textit{\textbf{M}} & \textit{\textbf{C}} \\ \hline
\textbf{\tiny{Cornell(CO)}} & 195 & 283 & 1,588 & 5 & \textbf{\tiny{Gplus(GP)}} & 700 & 28,055 & 887 & 4 \\
\textbf{\tiny{Texas(TE)}} & 185 & 280 & 1,501 & 5 & \textbf{\tiny{Cora}} & 2,708 & 5,278 & 1,432 & 7 \\
\textbf{\tiny{Washington(WA)}} & 217 & 366 & 1,578 & 5 & \textbf{\tiny{Citeseer(Cite)}} & 3,264 & 4,598 & 3,703 & 6 \\
\textbf{\tiny{Wisconsin(WI)}} & 262 & 459 & 1,623 & 5 & \textbf{\tiny{UAI2010(UAI)}} & 3,061 & 28,308 & 4,973 & 19 \\
\textbf{\tiny{Twitter(TW)}} & 155 & 3,442 & 1,470 & 7 & \textbf{\tiny{BlogCatalog(BL)}} & 5,196 & 171,743 & 8,189 & 6 \\
\textbf{\tiny{Facebook(FA)}} & 475 & 10,066 & 507 & 9 & \textbf{\tiny{Flikr(FL)}} & 7,575 & 239,738 & 12,047 & 9 \\ \hline
\end{tabular}
\end{table}

\textit{Cornell} (CO), \textit{Texas} (TE), \textit{Washington} (WA) and \textit{Wisconsin} (WI) are four sub-networks of the WebKB dataset\footnote{\scriptsize{http://www.cs.cmu.edu/afs/cs/project/theo-20/www/data/}}, which contains the hyperlinks and content of the web pages collected from the computer science departments in four universities. \textit{Twitter} (TW), \textit{Facebook} (FA) and \textit{Gplus} (GP) are the subsets of attributed ego-networks extracted from the Twitter\footnote{\scriptsize{http://snap.stanford.edu/data/ego-Twitter.html}}, Facebook\footnote{\scriptsize{http://snap.stanford.edu/data/ego-Facebook.html}} and Google+\footnote{\scriptsize{http://snap.stanford.edu/data/ego-Gplus.html}} datasets in Stanford Network Analysis Project (SNAP). \textit{Cora}\footnote{\scriptsize{http://www.cs.umd.edu/~sen/lbc-proj/data/cora.tgz}} \cite{Sen2008Collective} and \textit{Citeseer}\footnote{\scriptsize{http://www.cs.umd.edu/~sen/lbc-proj/data/citeseer.tgz}} (Cite) \cite{Sen2008Collective} are two science publication networks with the citation relations and paper content, while \textit{UAI2010} (UAI) \cite{Sen2008Collective} is a Wikipedia article citation network
including the reference relations and feature lists. \textit{BlogCatalog}\footnote{\scriptsize{http://github.com/xhuang31/AANE\_MATLAB/blob/master/BlogCatalog.mat}} (BL) \cite{Huang2017Accelerated} is a dataset collected from the blogger community BlogCatalog\footnote{\scriptsize{http://www.blogcatalog.com}} containing the interactive relations and interest tags of users. \textit{Flickr}\footnote{\scriptsize{http://github.com/xhuang31/AANE\_MATLAB/blob/master/Flickr.mat}} (FL) \cite{Huang2017Accelerated} is a social network of the online photo sharing platform Flickr\footnote{\scriptsize{https://www.flickr.com}}, which includes the friend relations between users and photos tags of each node.

\textbf{Baselines.} We adopted 11 network embedding approaches as baselines, which can be classified into three types. First, \textit{DeepWalk}\footnote{\scriptsize{https://github.com/phanein/deepwalk}} (DW) \cite{Perozzi2014DeepWalk}, \textit{node2vec}\footnote{\scriptsize{https://github.com/aditya\-grover/node2vec}} (N2V) \cite{Aditya2016node2vec}, \textit{LINE}\footnote{\scriptsize{https://github.com/tangjianpku/LINE}} \cite{Tang2015LINE}, \textit{SDNE}\footnote{\scriptsize{https://github.com/shenweichen/GraphEmbedding}} \cite{Wang2016Structural}, \textit{GraRep}\footnote{\scriptsize{https://github.com/ShelsonCao/GraRep}} \cite{Cao2015GraRep} and \textit{AROPE}\footnote{\scriptsize{https://github.com/ZW-ZHANG/AROPE}} \cite{Zhang2018Arbitrary-order} are methods exploring the high-order proximity of graph topology. Second, \textit{DNR} \cite{Liang2016Modularity} and \textit{M-NMF}\footnote{\scriptsize{https://github.com/thumanlab/M-NMF}} \cite{Wang2017Community} are approaches integrating the community structures. Moreover, \textit{TADW}\footnote{\scriptsize{https://github.com/albertyang33/TADW}} \cite{Cheng2015Network}, \textit{AANE}\footnote{\scriptsize{https://github.com/xhuang31/AANE\_MATLAB}} \cite{Huang2017Accelerated} and \textit{FSCNMF}\footnote{\scriptsize{https://github.com/benedekrozemberczki/FSCNMF}} (FSC) \cite{Bandy2018FSCNMF} are the embedding methods incorporating both graph topology and attributes.

To ensure the fairness of comparison, we set the embedding dimensionality $k=64$ for all the methods. Furthermore, we utilized the official open-source implementation of each baseline and adopted its default parameter setting.

For SGR, we adopted \textit{community structure} and \textit{attribute similarity} (represented by ${\bf{Q}}$ and ${\bf{S}}$, respectively) as the available side information (see Section~\ref{Sec:Meth}). Moreover, we utilized motifs $\{{M_0},{M_1},{M_2}\}$ (see Fig.~\ref{Motif-Fig}) to formulate the heterogeneous relations ${E_2}=\{({v_i},{a_w})\left| {{v_i} \in V,{a_w} \in A} \right.\}$. To effectively illustrate the effect of the hyper-parameters (i.e, $\{{\delta _1},{\delta _2},{\delta _3}\}$, $\{{\lambda _1},{\lambda _2}\}$) and the side-enhancement effect, we respectively recorded the evaluation metrics with the (\romannumeral1) default parameter setting (i.e., ${\delta _0}{\rm{ = }}{\delta _1}{\rm{ = }}{\delta _2}{\rm{ = }}1$), (\romannumeral2) fined-tuned parameters by adjusting ${\delta _0},{\delta _1},{\delta _2} \in \{ 0,1\}$ and (\romannumeral3) side-enhancement by adjusting ${\lambda _1},{\lambda _2} \in \{ 0,1\}$. The corresponding results are denoted as SGR(0), SGR(1) and SGR(R).

\textbf{Quantitative Evaluation.} In the evaluation, we adopted node clustering (a.k.a. community detection) and node classification as the testing downstream applications.

For node clustering, we applied the $K$Means algorithm to embeddings learned by all the methods and utilized the normalized mutual information (NMI) \cite{Cao2018Incorporating} as well as accuracy (AC) \cite{Cao2018Incorporating} as quality metrics. Moreover, we set the number of clusters in $K$Means according to the ground-truth of each dataset. For node classification, we utilized the support vector machine (SVM) implemented by the LibLinear package\footnote{\scriptsize{https://www.csie.ntu.edu.tw/\~cjlin/liblinear}} \cite{Fan2008LIBLINEAR} as the downstream classifier. For each dataset, we randomly select 10\% of the nodes to train the classifier, with the rest nodes as the test data. Accuracy (AC) \cite{Cui2017A} and Macro-F1 \cite{Cui2017A} were adopted as quality metrics.

Both the clustering and classification procedures were repeated $100$ times for each method and dataset. The average results in terms of NMI, (clustering) AC, (classification) AC, and Macro-F1 are shown in Table~\ref{Clus-NMI-Table}, \ref{Clus-AC-Table}, \ref{Clas-AC-Table} and \ref{Clas-F1-Table}, where the best metric is in \textbf{bold}; the second-best result is \underline{underlined}; '-' denotes there is no improvement for the parameter tuning SGR(1) or side-enhancement SGR(R) compared with the basic version SGR(0).

\begin{table}[t]\tiny
\caption{Evaluation result of node clustering in terms of NMI(\%)}\label{Clus-NMI-Table}
\centering
\begin{tabular}{p{0.55cm}|p{0.2cm}p{0.2cm}p{0.2cm}p{0.2cm}p{0.2cm}p{0.2cm}p{0.2cm}p{0.2cm}p{0.2cm}p{0.2cm}p{0.2cm}p{0.3cm}}
\hline
 & \textbf{CO} & \textbf{TE} & \textbf{WA} & \textbf{WI} & \textbf{TW} & \textbf{FA} & \textbf{GP} & \textbf{Cora} & \textbf{Cite} & \textbf{UAI} & \textbf{BL} & \textbf{FL} \\ \hline
DW & 6.79 & 5.79 & 7.22 & 7.41 & 33.98 & \underline{58.37} & 32.38 & 36.91 & 14.20 & 33.38 & 19.22 & 16.64 \\
N2V & 6.56 & 5.55 & 6.13 & 6.81 & 32.34 & 56.49 & 33.12 & 40.67 & 21.03 & 34.31 & 20.42 & 17.27 \\
LINE & 12.24 & 18.56 & 21.19 & 10.79 & 35.18 & 42.84 & 34.97 & 25.08 & 10.80 & 12.12 & 4.10 & 0.65 \\
SDNE & 13.78 & 16.85 & 24.42 & 8.98 & 28.02 & 28.22 & 27.73 & 10.96 & 4.04 & 11.44 & 9.62 & 3.96 \\
GraRep & 8.64 & 12.12 & 5.03 & 8.41 & 34.34 & 53.80 & 38.92 & 36.66 & 11.54 & 33.83 & 22.08 & 16.39 \\
AROPE & 9.08 & 10.29 & 9.63 & 6.36 & 29.32 & 25.42 & 19.45 & 8.85 & 4.54 & 13.63 & 14.37 & 8.40 \\
DNR & 11.91 & 18.70 & 24.29 & 9.46 & 32.45 & 31.29 & 25.47 & 16.09 & 6.48 & 5.66 & 13.28 & 4.42 \\
MNMF & 12.90 & 18.07 & 22.88 & 8.94 & 32.86 & 40.53 & \underline{40.38} & 10.34 & 5.18 & 19.36 & 17.25 & 14.76 \\
TADW & 11.49 & 8.45 & 10.67 & 13.15 & 25.34 & 46.34 & 5.56 & 29.71 & 19.29 & 25.33 & 7.85 & 2.27 \\
AANE & 30.63 & 30.32 & 38.11 & 40.91 & 33.72 & 49.76 & 37.72 & 17.78 & 19.82 & 41.18 & 28.26 & \underline{39.30} \\
FSC & 10.67 & 14.72 & 11.25 & 15.06 & 9.94 & 29.71 & 25.69 & 13.89 & 18.83 & 44.24 & 1.46 & 0.37 \\ \hline
SGR(0) & 31.67 & 33.97 & 40.23 & 40.29 & \underline{36.75} & \textbf{61.45} & 20.01 & \underline{49.33} & \textbf{39.62} & \underline{47.70} & \underline{31.87} & 20.66 \\
SGR(1) & \underline{35.59} & \underline{36.57} & \underline{45.19} & \underline{41.63} & \textbf{37.06} & - & - & \textbf{50.09} & - & \textbf{47.97} & \textbf{32.80} & 20.47 \\
SGR(R) & \textbf{36.80} & \textbf{38.59} & \textbf{46.78} & \textbf{45.72} & 31.93 & - & \textbf{54.35} & - & \underline{38.48} & - & - & \textbf{55.20} \\ \hline
\end{tabular}
\end{table}

%+---------------------+

\begin{table}[t]\tiny
\caption{Evaluation result of node clustering in terms of AC(\%)}\label{Clus-AC-Table}
\centering
\begin{tabular}{p{0.55cm}|p{0.2cm}p{0.2cm}p{0.2cm}p{0.2cm}p{0.2cm}p{0.2cm}p{0.2cm}p{0.2cm}p{0.2cm}p{0.2cm}p{0.2cm}p{0.3cm}}
\hline
 & \textbf{CO} & \textbf{TE} & \textbf{WA} & \textbf{WI} & \textbf{TW} & \textbf{FA} & \textbf{GP} & \textbf{Cora} & \textbf{Cite} & \textbf{UAI} & \textbf{BL} & \textbf{FL} \\ \hline
DW & 38.31 & 50.37 & 44.60 & 43.65 & 41.89 & 68.53 & 56.96 & 51.55 & 40.79 & 36.49 & 35.44 & 30.87 \\
N2V & 37.50 & 47.02 & 40.97 & 38.92 & 36.74 & 57.93 & 54.53 & 54.95 & 44.00 & 37.69 & 36.81 & 31.42 \\
LINE & 38.98 & 54.79 & 56.16 & 43.74 & 42.36 & 37.58 & 56.58 & 42.37 & 26.96 & 16.49 & 25.24 & 13.09 \\
SDNE & 42.12 & 54.95 & 62.35 & 47.43 & 36.06 & 26.83 & 55.27 & 31.25 & 22.51 & 19.17 & 26.88 & 15.52 \\
GraRep & 32.25 & 35.39 & 31.25 & 33.21 & 44.79 & 51.90 & 53.18 & 50.43 & 33.17 & 37.47 & 38.79 & 29.18 \\
AROPE & 42.95 & 56.02 & 48.91 & 46.51 & 37.63 & 27.58 & 55.87 & 32.68 & 23.05 & 21.24 & 28.18 & 18.27 \\
DNR & 37.59 & 52.42 & 55.34 & 42.71 & 44.57 & 31.57 & 53.31 & 33.56 & 23.67 & 12.89 & 32.57 & 18.52 \\
MNMF & 39.26 & 54.82 & 60.19 & 45.98 & 40.02 & 35.81 & 54.28 & 32.44 & 23.30 & 24.27 & 33.90 & 28.35 \\
TADW & 47.79 & 57.63 & 50.71 & 50.34 & 43.81 & 56.68 & 43.37 & 46.32 & 38.26 & 28.01 & 23.26 & 14.15 \\
AANE & 51.49 & 53.76 & 52.76 & 59.23 & 43.48 & \underline{69.88} & \underline{65.34} & 36.44 & 43.07 & 40.72 & \textbf{45.14} & \underline{38.57} \\
FSC & 45.73 & 58.41 & 50.62 & 51.95 & 35.52 & 41.37 & 60.55 & 35.14 & 41.38 & 41.15 & 18.92 & 11.81 \\ \hline
SGR(0) & 54.56 & 59.16 & 66.00 & \underline{64.87} & 45.41 & \textbf{71.37} & 54.64 & \underline{60.92} & \textbf{62.78} & \underline{45.61} & 41.85 & 33.56 \\
SGR(1) & \underline{56.25} & \textbf{63.50} & \underline{66.65} & 60.32 & \underline{46.95} & - & - & \textbf{62.44} & - & \textbf{46.04} & \underline{43.10} & 33.43 \\
SGR(R) & \textbf{59.11} & \underline{60.97} & \textbf{71.41} & \textbf{70.85} & \textbf{52.68} & - & \textbf{83.22} & - & \underline{59.92} & - & - & \textbf{61.91} \\ \hline
\end{tabular}
\end{table}

%+---------------------+

\begin{table}[t]\tiny
\caption{Evaluation result of node classification in terms of AC(\%)}\label{Clas-AC-Table}
\centering
\begin{tabular}{p{0.55cm}|p{0.2cm}p{0.2cm}p{0.2cm}p{0.2cm}p{0.2cm}p{0.2cm}p{0.2cm}p{0.2cm}p{0.2cm}p{0.2cm}p{0.2cm}p{0.3cm}}
\hline
 & \textbf{CO} & \textbf{TE} & \textbf{WA} & \textbf{WI} & \textbf{TW} & \textbf{FA} & \textbf{GP} & \textbf{Cora} & \textbf{Cite} & \textbf{UAI} & \textbf{BL} & \textbf{FL} \\ \hline
DW & 32.31 & 47.34 & 38.85 & 40.66 & 48.42 & 84.44 & 85.91 & 68.24 & 45.32 & 45.71 & 59.63 & 44.54 \\
N2V & 31.85 & 47.40 & 39.83 & 40.60 & 48.54 & \underline{85.38} & 87.10 & 72.53 & 50.18 & 48.50 & 59.94 & 45.99 \\
LINE & 37.19 & 57.39 & 53.07 & 48.64 & 46.88 & 83.56 & 90.08 & 71.06 & 44.90 & 35.05 & 32.35 & 12.39 \\
SDNE & 38.01 & 56.62 & 59.36 & 48.25 & 45.22 & 75.04 & 89.32 & 38.66 & 23.35 & 26.55 & 56.37 & 31.73 \\
GraRep & 36.00 & 50.99 & 41.74 & 46.03 & \textbf{51.12} & 84.98 & 88.45 & 73.96 & 48.75 & 52.61 & 65.87 & 50.24 \\
AROPE & 37.69 & 55.37 & 50.50 & 46.30 & 48.67 & 81.78 & \underline{93.90} & 65.14 & 43.13 & 45.42 & 67.12 & 57.17 \\
DNR & 38.46 & 57.37 & 58.63 & 50.24 & 46.88 & 72.09 & 91.88 & 44.49 & 27.31 & 17.90 & 40.79 & 17.80 \\
MNMF & 37.94 & 57.85 & 46.86 & 46.67 & 50.35 & 83.56 & 91.03 & 69.57 & 46.94 & 45.26 & 66.45 & 53.29 \\
TADW & 45.07 & 52.14 & 48.73 & 50.42 & 46.74 & 79.57 & 83.77 & 67.26 & 56.89 & 55.43 & 89.76 & 56.65 \\
AANE & \textbf{60.91} & 65.01 & 69.33 & \underline{72.34} & 37.18 & 71.55 & 84.85 & 72.62 & 65.60 & 61.29 & 82.19 & \underline{85.97} \\
FSC & 54.14 & \underline{65.84} & 64.14 & 68.11 & 41.63 & 72.45 & 81.85 & 61.01 & 62.57 & 69.86 & 73.11 & 49.45 \\ \hline
SGR(0) & 55.61 & 63.89 & 67.11 & 69.26 & 48.69 & \textbf{85.76} & \textbf{94.74} & \underline{80.03} & \underline{67.62} & \underline{70.72} & \underline{90.11} & 84.25 \\
SGR(1) & 57.36 & 65.07 & \textbf{71.20} & 72.10 & 50.73 & - & - & \textbf{80.29} & - & \textbf{70.91} & \textbf{90.18} & 84.36 \\
SGR(R) & \underline{58.82} & \textbf{71.17} & \underline{69.64} & \textbf{74.14} & \underline{50.82} & - & 90.17 & - & \textbf{71.31} & - & - & \textbf{89.02} \\ \hline
\end{tabular}
\end{table}

%+--------------------+

\begin{table}[t]\tiny
\caption{Evaluation result of node classification in terms of F1(\%)}\label{Clas-F1-Table}
\centering
\begin{tabular}{p{0.55cm}|p{0.2cm}p{0.2cm}p{0.2cm}p{0.2cm}p{0.2cm}p{0.2cm}p{0.2cm}p{0.2cm}p{0.2cm}p{0.2cm}p{0.2cm}p{0.3cm}}
\hline
 & \textbf{CO} & \textbf{TE} & \textbf{WA} & \textbf{WI} & \textbf{TW} & \textbf{FA} & \textbf{GP} & \textbf{Cora} & \textbf{Cite} & \textbf{UAI} & \textbf{BL} & \textbf{FL} \\ \hline
DW & 20.05 & 21.15 & 20.79 & 24.22 & 25.87 & 54.86 & 53.69 & 66.86 & 41.95 & 37.68 & 59.11 & 43.61 \\
N2V & 19.55 & 20.32 & 21.03 & 24.61 & 18.29 & 54.46 & 51.44 & 70.82 & 45.96 & 39.42 & 59.30 & 44.71 \\
LINE & 24.10 & 27.36 & 28.17 & 28.72 & 25.18 & 54.19 & 55.85 & 69.20 & 40.51 & 24.33 & 27.89 & 10.42 \\
SDNE & 22.63 & 25.69 & 30.08 & 26.02 & 24.56 & 44.88 & 64.81 & 22.41 & 10.32 & 20.18 & 55.63 & 27.59 \\
GraRep & 25.48 & 29.50 & 24.48 & 28.76 & \textbf{27.41} & \underline{54.97} & 59.48 & 72.76 & 45.20 & 43.10 & 65.22 & 49.73 \\
AROPE & 24.44 & 28.89 & 26.89 & 28.33 & 25.74 & 46.28 & \underline{71.52} & 63.47 & 39.36 & 35.09 & 65.79 & 55.58 \\
DNR & 19.18 & 25.70 & 28.32 & 22.59 & 24.57 & 30.01 & 63.09 & 29.26 & 17.28 & 7.71 & 39.61 & 13.40 \\
MNMF & 23.20 & 30.50 & 24.77 & 27.94 & 27.10 & 54.89 & 65.39 & 67.20 & 42.81 & 35.73 & 65.54 & 52.46 \\
TADW & 29.02 & 21.89 & 27.00 & 29.56 & 26.60 & 45.90 & 50.15 & 65.02 & 52.79 & 43.87 & 89.59 & 55.90 \\
AANE & 39.95 & 31.50 & 36.68 & 44.02 & 11.29 & 26.53 & 43.91 & 69.16 & 59.12 & 40.99 & 81.75 & \underline{85.63} \\
FSC & 36.18 & 36.01 & 38.78 & 44.64 & 20.67 & 28.58 & 42.86 & 58.29 & 55.95 & 55.21 & 72.61 & 46.16 \\ \hline
SGR(0) & 38.58 & 37.77 & 40.33 & 43.41 & 25.13 & \textbf{57.83} & \textbf{73.38} & \underline{78.54} & \underline{62.52} & \underline{57.85} & \underline{89.96} & 84.06 \\
SGR(1) & \underline{40.08} & \underline{38.40} & \underline{42.42} & \underline{46.94} & 25.99 & - & - & \textbf{78.74} & - & \textbf{58.24} & \textbf{90.04} & 84.16 \\
SGR(R) & \textbf{44.34} & \textbf{45.86} & \textbf{44.43} & \textbf{52.78} & \underline{27.12} & - & 66.25 & - & \textbf{63.05} & - & - & \textbf{88.64} \\ \hline
\end{tabular}
\end{table}

In node clustering, SGR (including SGR(0), SGR(1) and SGR(R)) has the best performance on all the 12 datasets in terms of NMI and outperforms other baselines on 11 datasets in terms of AC, with average improvement of \textbf{25.31}\% and \textbf{21.22}\% w.r.t. second-best baselines. In node classification, SGR achieves the best performance on 10 and 11 datasets for metrics of AC and Macro-F1, with average improvement of \textbf{3.74}\% and \textbf{9.39}\% w.r.t. second-best baselines.

In comparison with SGR(0) and SGR(1), the performance of node clustering and classification can be further improved on most datasets via the side-enhancement (i.e., SGR(R)) but the improvement cannot be ensured for all the datasets (e.g., \textit{Facebook}, \textit{UAI2010}, \textit{Cora} and \textit{BlogCatalog}). A possible reason is that the incorporation of side information (e.g., community structure) may not only bring complementary knowledge of the graph but also introduce inconsistent features or noise into the learned embeddings, affecting the performance of downstream tasks \cite{qin2018adaptive,qin2021dual}.

\begin{figure}[]\footnotesize
\centering
\begin{minipage}{0.48\linewidth}
    \centering
    \subfigure[\scriptsize{Cornell, SGR(0) ($o$)}]
    {\includegraphics[width=1.0\textwidth, trim=5 0 5 0, clip]{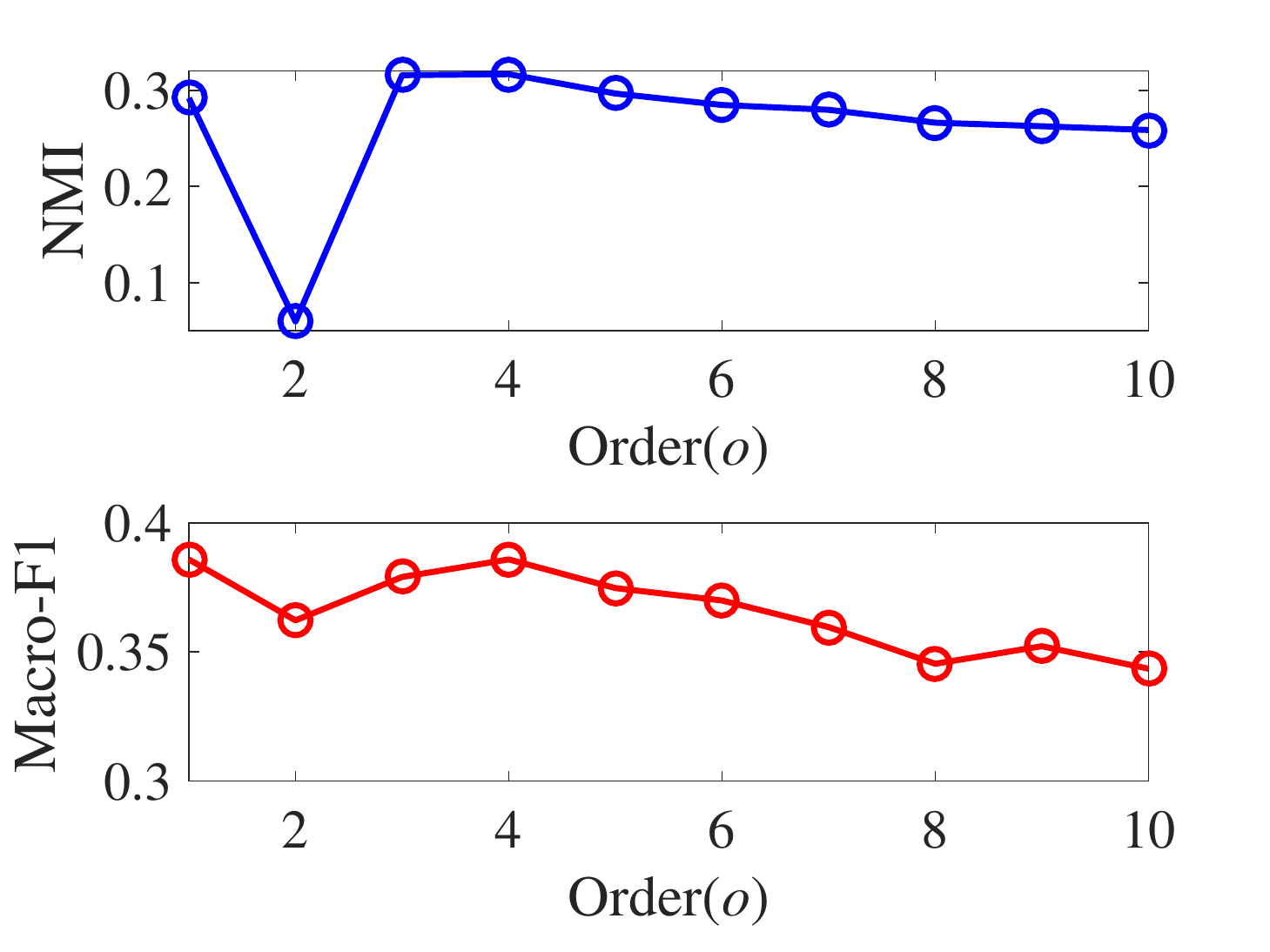}}
\end{minipage}
\begin{minipage}{0.48\linewidth}
    \centering
    \subfigure[\scriptsize{Cornell, SGR(1) (${\delta_*}$)}]
    {\includegraphics[width=1.0\textwidth, trim=5 0 5 0, clip]{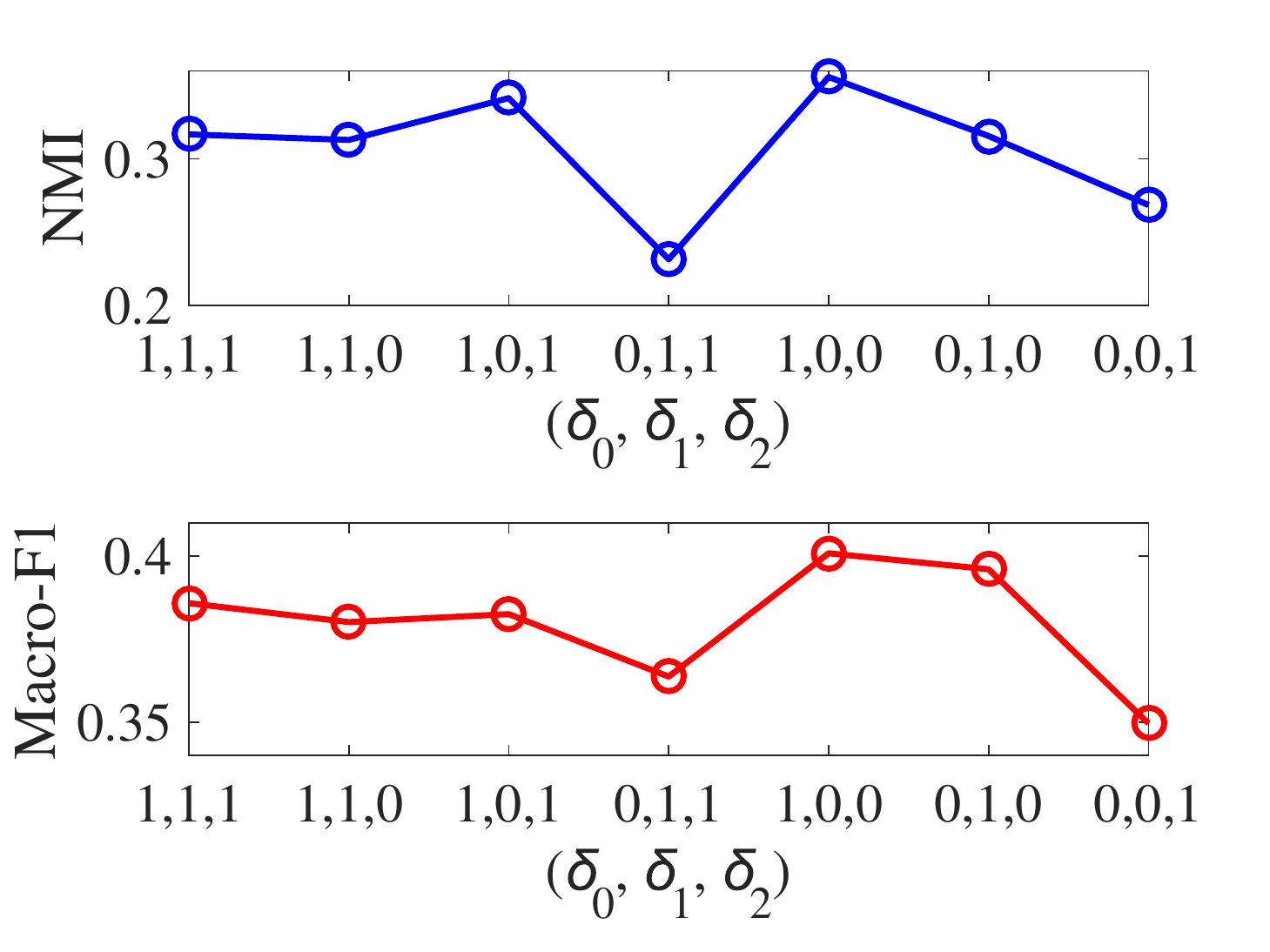}}
\end{minipage}
\begin{minipage}{0.48\linewidth}
    \centering
    \subfigure[\scriptsize{Cornell, SGR(R) (${\lambda_*}$)}]
    {\includegraphics[width=1.0\textwidth, trim=5 0 5 0, clip]{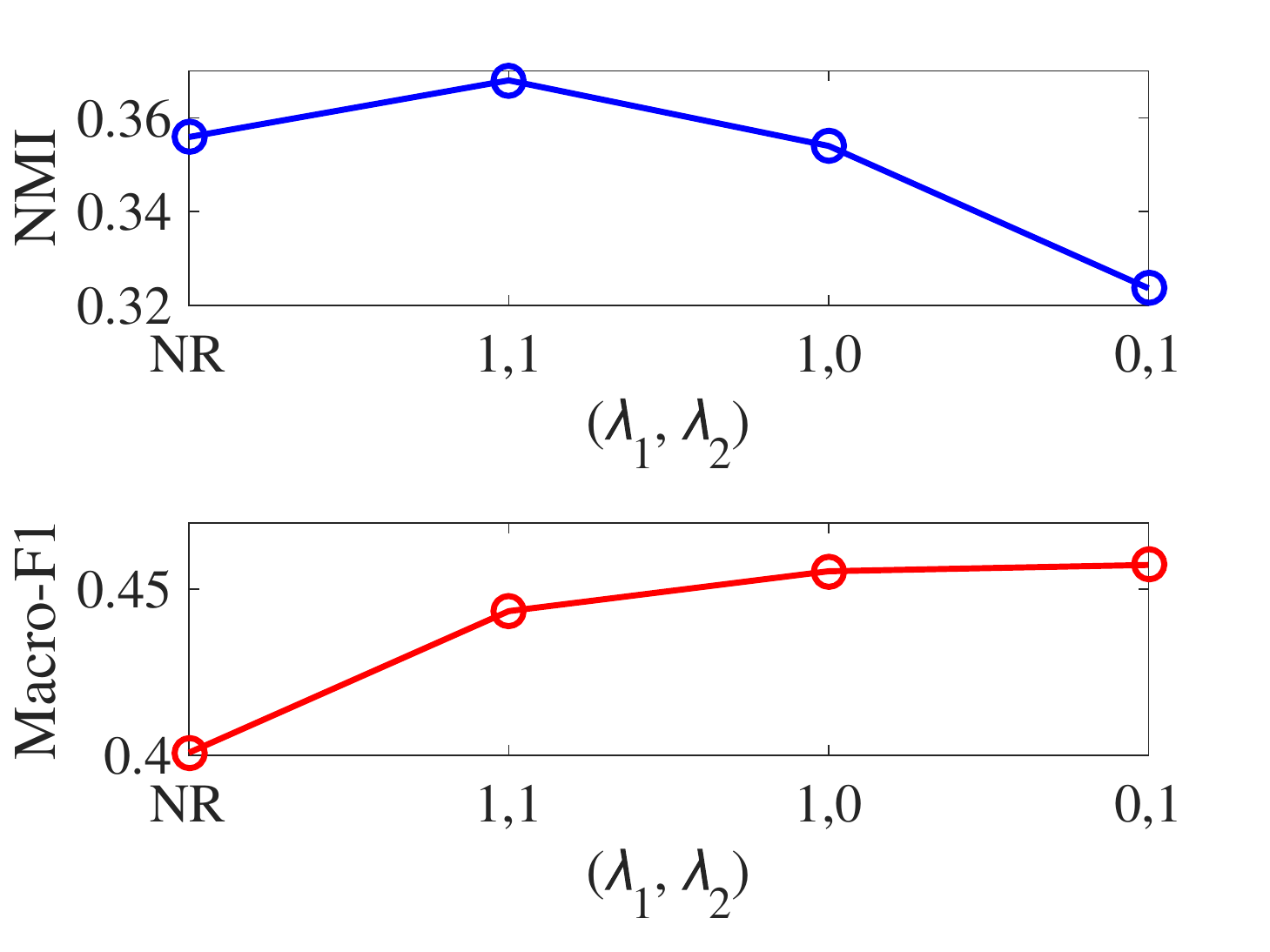}}
\end{minipage}
\begin{minipage}{0.48\linewidth}
    \centering
    \subfigure[\scriptsize{Citeseer, SGR(0) ($o$)}]
    {\includegraphics[width=1.0\textwidth, trim=5 0 5 0, clip]{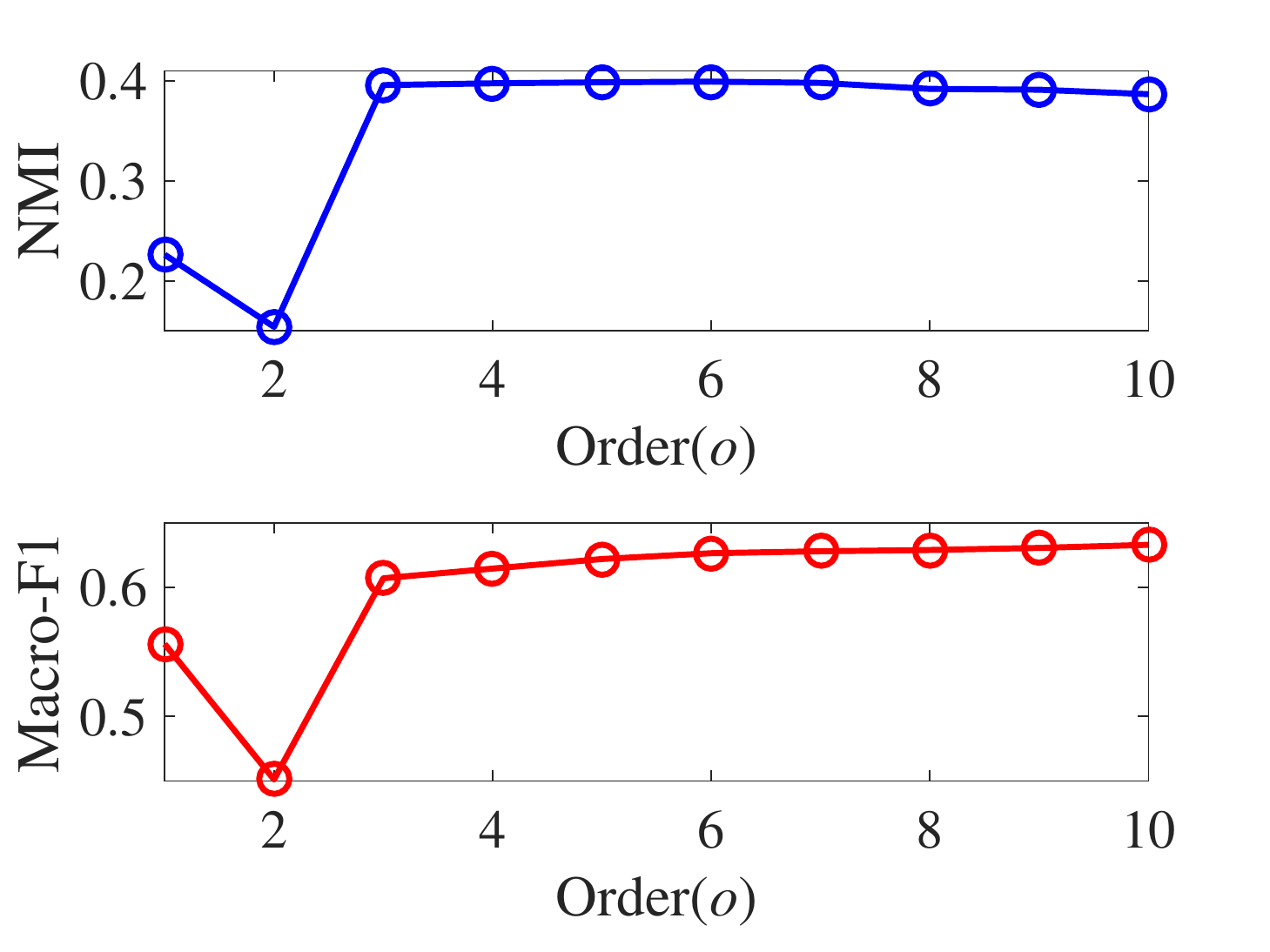}}
\end{minipage}
\begin{minipage}{0.48\linewidth}
    \centering
    \subfigure[\scriptsize{Citeseer, SGR(1) (${\delta_*}$)}]
    {\includegraphics[width=1.0\textwidth, trim=5 0 5 0, clip]{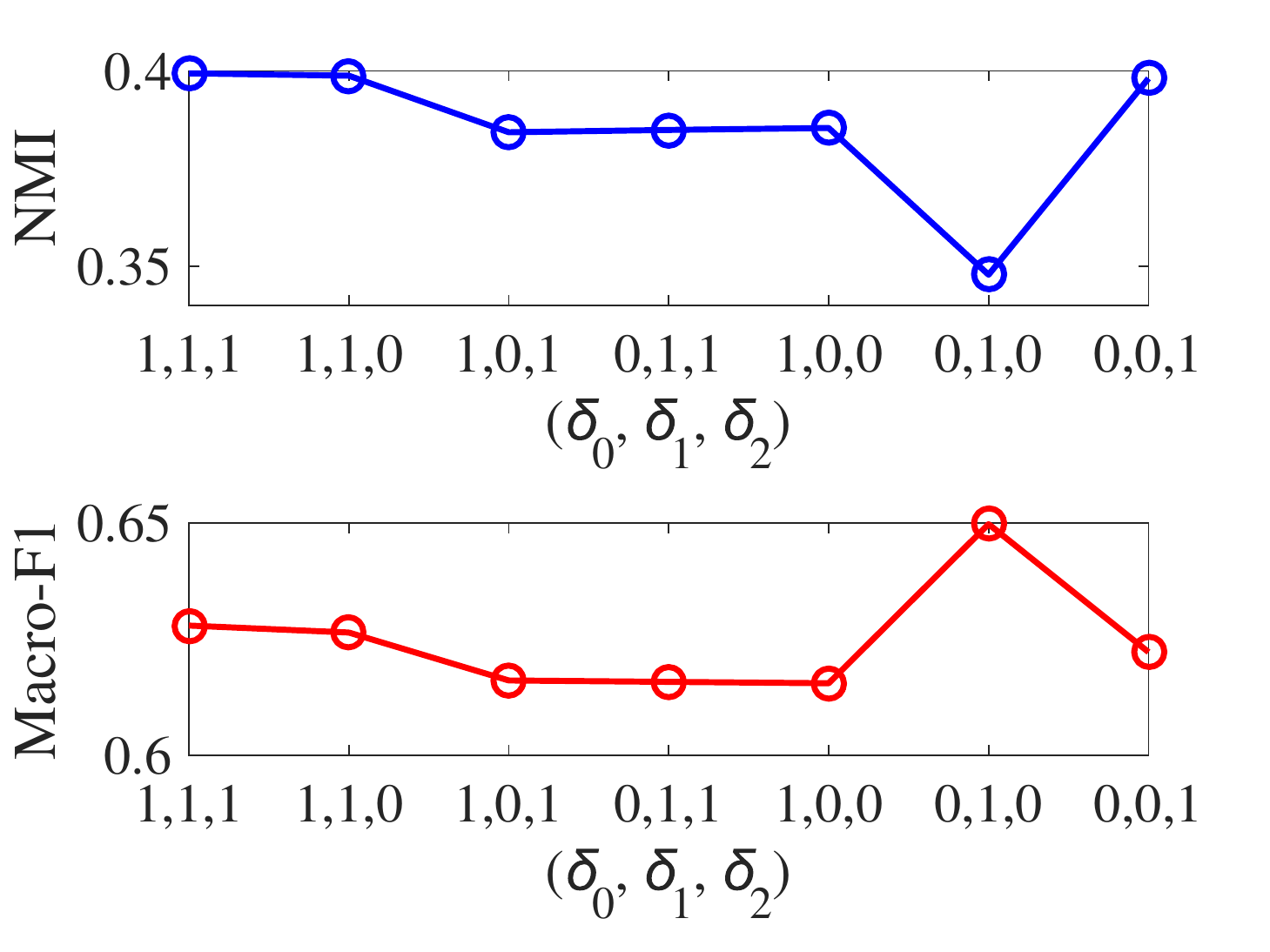}}
\end{minipage}
\begin{minipage}{0.48\linewidth}
    \centering
    \subfigure[\scriptsize{Citeseer, SGR(R) (${\lambda_*}$)}]
    {\includegraphics[width=1.0\textwidth, trim=5 0 5 0, clip]{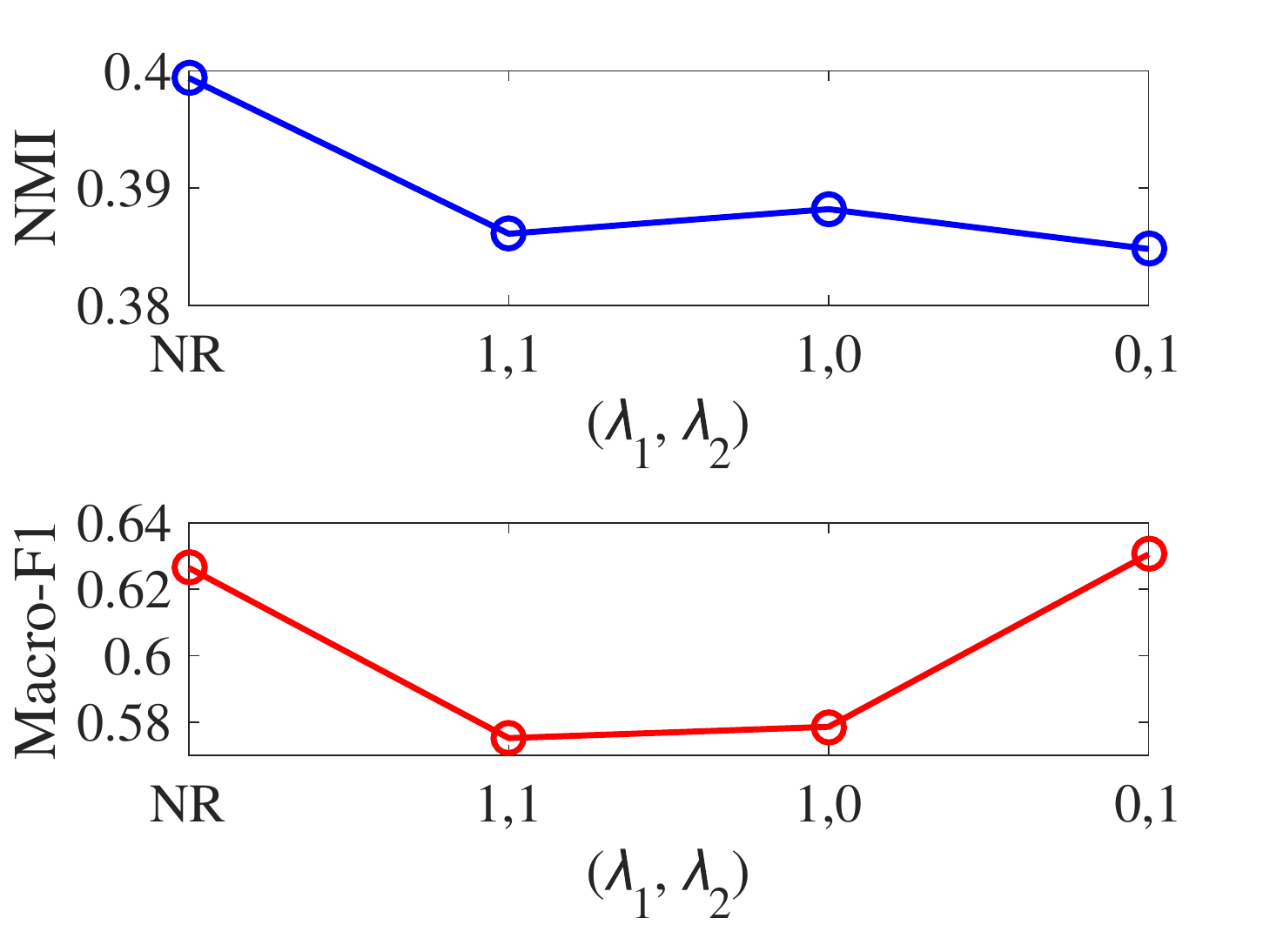}}
\end{minipage}
\caption{Parametric Analysis on \textit{Cornell} and  \textit{Citeseer}} \label{Param-Fig}
\end{figure}

In the experiments, we first fixed ${\delta_1} = {\delta_2} = {\delta_3} = 1$ and adjusted the proximity order $o \in \{ 1,2, \cdots ,10\}$ for each dataset, with the best metric reported for SGR(0). Based on the selected setting of $o$, we tuned ${\delta _0},{\delta _1},{\delta _2} \in \{ 0,1\}$ for SGR(1). Moreover, the side-information is further incorporated (based on the parameter setting of SGR(0) and SGR(1)) by adjusting ${\lambda _1},{\lambda _2} \in \{ 0,1\}$ for SGR(R). Due to the space limit, we demonstrate the parameter adjustment (in terms of NMI and Macro-F1 for node clustering and classification) of \textit{Cornell} and \textit{Citeseer} as 2 examples in Fig.~\ref{Param-Fig}, where 'NR' represents the baseline performance of SGR(1) (compared with SGR(R)).

According to Fig.~\ref{Param-Fig} and our records, different settings of $o$ may result in significantly different pefromance for a certain dataset, indicating that the order of random walk on $G'$ is a primary factor that affects the incorporation of graph topology and attributes. Moreover, the best performance in one task (e.g., node clustering or classification) does not imply the best result in other tasks w.r.t. a certain parameter setting (e.g., $({\delta _0},{\delta _1},{\delta _2}) = (0,1,0)$ for \textit{Citeseer} and $({\lambda _1},{\lambda _2}) = (1,1)$ for \textit{Cornell}). For each dataset, we determine the best parameter setting by comprehensively considering the performance of both node clustering and classification. Although it is hard to find a fixed parameter setting that can ensure the best performance for all the datasets, we recommend setting $o=4$ and ${\delta _0}{\rm{ = }}{\delta _1}{\rm{ = }}{\delta _2}{\rm{ = }}1$, in which SGR can achieve a relatively high performance (despite not the best) for most datasets. Better results can be obtained by fine-tuning ${\delta _0},{\delta _1},{\delta _2} \in \{ 0,1\}$ and utilizing the side-enhancement (i.e., adjust ${\lambda _1},{\lambda _2} \in \{ 0,1\}$).

We used MATLAB to implement SGR. The average runtime in terms of second with and without the side-enhancement on a server (Intel Xeon CPU E5-2680 v4 @2.40GHz and 32GB main memory) is shown in Table~\ref{Time-Table}. To further speed up the computation, some distributed fast SVD approaches (e.g., \cite{Iwen2016A}) and optimized matrix operation libraries (e.g., OpenBLAS\footnote{\scriptsize{http://www.openblas.net}}) can be utilized.

\begin{table}[t]\tiny
\caption{Average Runtime (sec) of SGR}\label{Time-Table}
\centering
\begin{tabular}{p{0.55cm}|p{0.1cm}p{0.1cm}p{0.1cm}p{0.1cm}p{0.1cm}p{0.1cm}p{0.1cm}p{0.2cm}p{0.25cm}p{0.25cm}p{0.35cm}p{0.45cm}}
\hline
 & \textbf{CO} & \textbf{TE} & \textbf{WA} & \textbf{WI} & \textbf{TW} & \textbf{FA} & \textbf{GP} & \textbf{Cora} & \textbf{Cite} & \textbf{UAI} & \textbf{BL} & \textbf{FL} \\ \hline
\textbf{SGR} & 1.57 & 1.34 & 1.61 & 1.81 & 1.20 & 0.35 & 1.19 & 16.39 & 66.46 & 100.06 & 462.93 & 1259.32 \\
\textbf{SGR}(R) & 2.46 & 1.89 & 3.22 & 4.26 & 3.12 & 0.51 & 1.62 & 27.87 & 118.40 & 270.37 & 1146.97 & 9997.84 \\ \hline
\end{tabular}
\end{table}

\subsection{Case Study}
We utilized the \textit{LastFM} dataset\footnote{\scriptsize{https://grouplens.org/datasets/hetrec-2011}} \cite{Cantador:RecSys2011} to illustrate SGR's ability to generate the semantic descriptions and adopted semantic community detection \cite{Wang2016Semantic,qin2018adaptive,qin2021dual,li2019identifying} as the testing application, where we can generate one or more wordclouds for each community when the community partition is finished. The dataset was collected from the online music platform last.fm\footnote{\scriptsize{http://www.lastfm.com}}, including the friend relations and interest tags of the users. After pre-processing, the dataset contains $1,892$ users (nodes), $12,717$ friend relations (edges) and $9,749$ tags (attributes).

\begin{figure}[]
\centering
\begin{minipage}[b]{0.32\linewidth}
    \centering
    \subfigure[\scriptsize{Visualization of Node \& Attribute Embeddings}]
    {\includegraphics[width=1.0\textwidth, trim=5 0 5 0, clip]{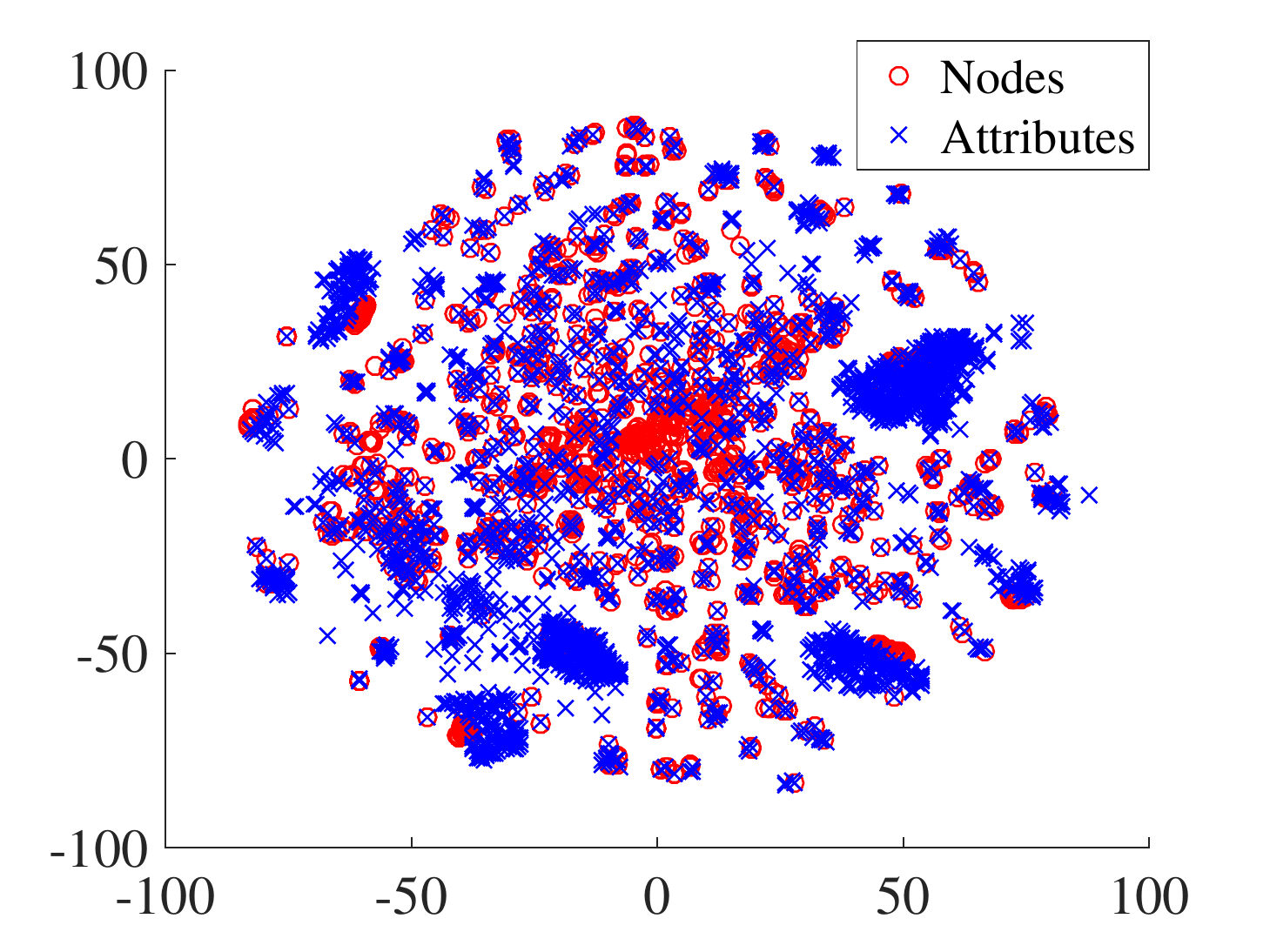}}
\end{minipage}
\begin{minipage}[b]{0.32\linewidth}
    \centering
    \subfigure[\scriptsize{Visualization of Cluster Centers}]
    {\includegraphics[width=1.0\textwidth, trim=5 0 5 0, clip]{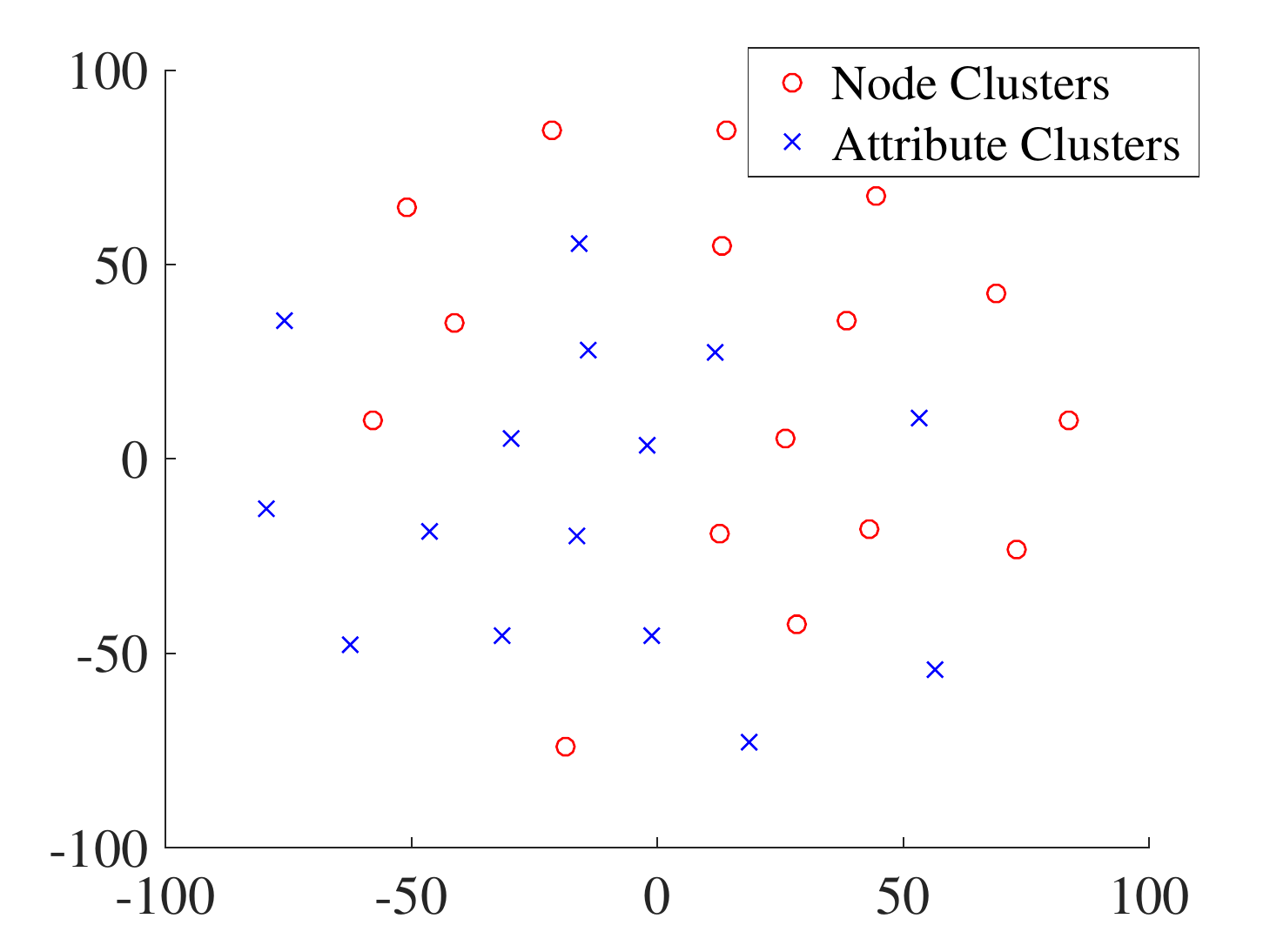}}
\end{minipage}
\begin{minipage}[b]{0.32\linewidth}
    \centering
    \subfigure[\scriptsize{Description of Community \#15}]
    {\includegraphics[width=1.0\textwidth, trim=0 0 0 0, clip]{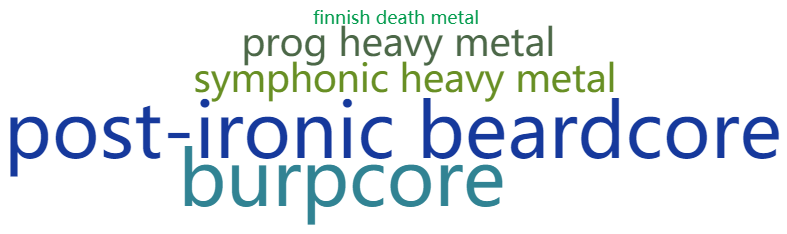}}
\end{minipage}
\begin{minipage}[b]{0.32\linewidth}
    \centering
    \subfigure[\scriptsize{Top-1 Description of Community \#1}]
    {\includegraphics[width=1.0\textwidth, trim=0 0 0 0, clip]{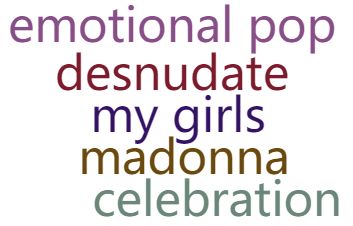}}
\end{minipage}
\begin{minipage}[b]{0.32\linewidth}
    \centering
    \subfigure[\scriptsize{Top-2 Description of Community \#1}]
    {\includegraphics[width=1.0\textwidth, trim=0 0 0 0, clip]{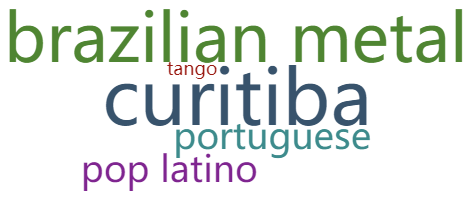}}
\end{minipage}
\begin{minipage}{0.32\linewidth}
    \centering
    \caption{Visualization Results of the Case Study on \textit{LastFM}}
	\label{lastFM_case-fig}
\end{minipage}
\end{figure}

We obtained the node embeddings and attribute embeddings (notated as $\{ {{{\bf{x'}}}_{{v_i}}}\}$ and $\{ {{{\bf{x'}}}_{{a_w}}}\}$) by setting $o=4$, ${\delta _0}{\rm{ = }}{\delta _1}{\rm{ = }}{\delta _2}{\rm{ = }}1$ for the basic version of SGR (i.e., SGR(0)). We further utilized t-SNE \cite{Der2008Visualizing} to map the representations $\{{{{\bf{x'}}}_{{v_i}}},{{{\bf{x'}}}_{{a_w}}}\}$ into a 2D space with the result visualized in Fig.~\ref{lastFM_case-fig} (a), where $\{{{\bf{x'}}_{{v_i}}},{{\bf{x'}}_{{a_w}}}\}$ are well mapped into a common hidden space, indicating the comparability of the heterogeneous embeddings. Namely, the distance between $({{{\bf{x'}}}_{{v_i}}},{{{\bf{x'}}}_{{a_w}}})$ can be used to measure the heterogeneous similarity between node ${v_i}$ and attribute ${a_w}$. Furthermore, we applied the advanced X-Means algorithm \cite{Ishioka2000Extended} to determine the proper number of clusters and the corresponding clustering membership respectively for $\{ {{{\bf{x'}}}_{{v_i}}}\}$ and $\{ {{{\bf{x'}}}_{{a_w}}}\}$. Finally, we set the number of topology clusters and attribute clusters (i.e., ${K_1}$ and ${K_2}$) to be 16 and 15. The cluster centers of both $\{{{{\bf{x'}}}_{{v_i}}}\}$ and $\{ {{{\bf{x'}}}_{{a_w}}}\}$ are also mapped into a 2D space via t-SNE. The corresponding visualization result is shown in Fig.~\ref{lastFM_case-fig} (b), where each community (i.e., node cluster) may have one or more nearest attribute cluster centers with close distance in the hidden space, indicating that it's possible for a community to have more than one relevant semantic topics.

We adopted two different strategies to generate the descriptions. First, we selected the top-5 relevant keywords for each community by calculating and ranking the distance between each node cluster center and attribute embeddings ${{{\bf{x'}}}_{{a_w}}}$, where each community only have one comprehensive description. The single wordcloud of community \#15 is illustrated in Fig.~\ref{lastFM_case-fig} (c) as an example. Second, we selected the most relevant topics for each community by measuring the distance between the cluster centers of node and attribute, and then generated the top-5 keywords for each topic. In this case, each community may have more than one descriptions, with each one reflecting a specific aspect of semantic (i.e., topic). As an instance, two relevant descriptions of community \#1 are presented in Fig.~\ref{lastFM_case-fig} (d) and (e).

To verify the semantic relations among the top words in each wordcloud, we used them as the query words of Wikipedia\footnote{\scriptsize{https://en.wikipedia.org/}} to refer to other related materials. In Fig.~\ref{lastFM_case-fig} (c), \textit{metal music} is the primary topic, just as `finnish death metal', `prog heavy metal', and `symphonic heavy metal' infer. `post-ironic beardcore' and `burpcore' are both words used to describe the genre of metalcore music, which is a fusion genre of hardcore punk and extreme metal. For Fig.~\ref{lastFM_case-fig} (d), \textit{Pop music} should be the hidden topic, where `emotional pop' is directly related to it. `desnudate' and `my girls' may refer to the songs of American pop singer Christina Aguilera in album Bionic, which is characterized with the genres of electropop and futurepop. Moreover, `madonna' may indicate the American singer Madonna Louise Ciccone, who is referred to as the `Queen of Pop'. The topic of Fig.~\ref{lastFM_case-fig} (d) is \textit{Latin music}, where `pop latino', `brazilian metal', and `portuguese' are related to the music and language in Latin America. Furthermore, `curitiba' is a city in Brazil while `tango' is a popular dance originated in Latin America.

\section{Conclusion}\label{Sec:Conc}
In this paper, we introduced a novel SGR method to generally formulate the network embedding in attributed graphs as a high-order proximity based embedding task of an auxilairy weighted graph with heterogeneous entities. The proposed model could not only comprehensively capture the high-order proximity inside and among graph structure and semantic, but also jointly learn the low-dimensional representations of both nodes and attributes, effectively supporting the advanced semantic-oriented downstream applications (e.g., semantic community detection). The effectiveness of SGR was also verified on a series of real attributed graphs for several graph inference tasks.

In our future work, we intend to explore a more comprehensive but simpler parameter adjustment strategy to effectively reduce the hyper-parameters' search space, with the guarantee of performance. Moreover, to further extend SGR to dynamic graphs \cite{lei2018adaptive,lei2019gcn,qin2022temporal,qin2023high}, exploring the non-linear high-order correlations among dynamic topology and attributes, is also our next research focus.


\bibliographystyle{IEEEtran}
%\bibliography{SGR-Ref}

\end{document}